\documentclass{article}
\usepackage[english]{babel}
\usepackage[font=small,labelfont=bf]{caption}
\usepackage[
  a4paper,
  top=2.5cm,
  bottom=2.5cm,
  left=2.3cm,
  right=2.3cm,
  heightrounded
]{geometry}

\usepackage[utf8]{inputenc}
\usepackage[T1]{fontenc}
\usepackage{appendix}
\usepackage{braket}
\usepackage{bm}
\usepackage{xcolor}
\usepackage{graphicx}
\usepackage{multirow}
\usepackage{amsmath,amssymb,amsfonts}
\usepackage{booktabs} 
\usepackage[normalem]{ulem}
\usepackage[colorlinks=true, allcolors=blue]{hyperref}

\newcommand{\eq}[1]{Eq.~\eqref{#1}}
\newcommand{\fig}[1]{Fig.~\ref{#1}}
\newcommand{\tab}[1]{Tab.~\ref{#1}}
\newcommand{\be}{\begin{equation}}
\newcommand{\ee}{\end{equation}}
\def\bea{\begin{eqnarray} }
\def\eea{ \end{eqnarray} } 

\newcommand{\Mpbar}{\overline{M}_{\text{Pl}}}
\newcommand{\Mp}{M_{\text{Pl}}}

\newcommand{\MG}{M_{\text{GUT}}}
\newcommand{\MEW}{M_{\text{EW}}}
\newcommand{\MR}{M_{R}}
\newcommand{\GeV}{\,\text{GeV}}

\newcommand{\rsw}{\rm sw}
\newcommand{\turb}{\rm turb}

\DeclareMathOperator{\Tr}{Tr}
\newcommand{\rep}[1]{\ensuremath{\mathbf{#1}}}

\begin{document}

\begin{titlepage}

\begin{flushright}
CQUeST-2025-0759
\end{flushright}
\bigskip

\begin{center}
{\Large\bfseries
Gravitational Wave Signals in a Promising Realization of
$\boldsymbol{SO(10)}$ Unification}

\bigskip
Injun Jeong$^{1,2}$,
Jörn Kersten$^{3,4}$,
Stefano Scopel$^{1,2}$,
and Liliana Velasco-Sevilla$^{1,2}$

\bigskip
\small
$^1$
Center for Quantum Spacetime, Sogang University,
35 Baekbeom-ro, Seoul 121-742, South Korea

$^2$
Department of Physics, Sogang University,
35 Baekbeom-ro, Seoul 121-742, South Korea

$^3$
Department of Physics and Technology, University of Bergen, Postboks 7803, 5020 Bergen, Norway

$^4$
Department of Physics and IPAP, Yonsei University,
50 Yonsei-ro, Seoul 03722, South Korea
\end{center}
\bigskip

\begin{abstract}
\noindent
We investigate gravitational wave signals in a non-supersymmetric grand unified model where the group $SO(10)$ is broken in two steps to the Standard Model gauge group.  We calculate the analytical form of the one-loop effective potential responsible for the first step of symmetry breaking and show that it can lead to a first-order phase transition with gravitational wave production. We also determine the gravitational wave background produced by the primordial plasma of relativistic particles.
The present experimental sensitivity is still far from the expected signals
but could be in reach of novel detector concepts.
\end{abstract}

\end{titlepage}

\section{Introduction}
Before the advent of gravitational wave (GW) experiments, physics beyond the Standard Model (SM) could be probed at early cosmological times using photons and neutrinos. The earliest observable photons come from the Cosmic Microwave
Background (CMB), because before recombination the Universe was optically thick, while detection of cosmic neutrinos could probe the primordial nucleosynthesis era. However, times at which Grand Unified Theories (GUTs) are unbroken remain inaccessible to cosmological information from photons and neutrinos.
On the other hand, the Large Hadron Collider has not found any
indication for physics beyond the SM, and hence observations from other experiments are crucial to keep probing compelling frameworks motivated by unsolved issues in the SM\@. 
As has been known for a long time \cite{Kamionkowski:1993fg}, GUTs provide one such framework that could leave a
signal in the form of a stochastic GW background. 

Due to the possible observation of a stochastic GW background
by pulsar timing arrays
\cite{NANOGrav:2023gor,EPTA:2023fyk,Reardon:2023gzh,Xu:2023wog} there has been a resurgence in studying GW from GUTs (see \cite{Caldwell:2022qsj} for extensive references).
The first evidence reported by the NANOGrav collaboration~\cite{Arzoumanian:2020vkk}
appeared to be consistent with the expected signal from cosmic strings,
considered the smoking gun of the characteristic sequence of symmetry
breakings expected in GUTs.
It now appears likely that what is observed by the pulsar timing arrays
cannot be the signal of a Nambu-Goto string arising from a GUT
\cite{Kume:2024xve}.
Nevertheless, signals of this kind can appear below scales of order
$10^{15}\GeV$, owing to the fact that there could be different stages of
symmetry breaking in GUT models, both supersymmetric and
non-supersymmetric. In addition, specific GUT scenarios can be characterized by the appearance of a rich pattern of mutually interacting topological defects~\cite{Dunsky:2021tih,Chun:2021brv}.

The current, upcoming and projected experiments have the potential to
detect GW signals in an extended frequency range. This provides an
invaluable opportunity to analyze under which conditions these could
come from a GUT group, to which specific scenario they correspond and at
what scale they appear. 
In this work, we will focus on two types of
signals: the GW produced by the high-temperature first-order phase
transitions (FOPT) induced by the GUT symmetry breaking
(Section~\ref{sec:FOPT}) and the stochastic background produced by the
shear viscosity of the relativistic plasma in thermal equilibrium
throughout the expansion history of the Universe
(Section~\ref{sec:Plasma}). 

As far as the former signal is concerned, it is not present in the SM and is considered a smoking gun for GUTs.
Due to the complexity of the construction of such models, there have
been no studies on FOPT in detailed GUT models, particularly at the GUT
scale. It is clear that one reason for this is the high frequency
required for the observation of the signals, and hence the majority of
FOPT analyzed within the context of GUTs refer to phase transitions in
the different chain of breakings of GUT models in the LIGO or LISA
region \cite{Huang:2017laj,Croon:2018kqn,Brdar:2019fur,Graf:2021xku}. However, given the panacea that GUT models are for addressing issues that in the SM remain unsolved, the study of all possible signals that GUT theories can produce merits consideration and therefore we begin the task of studying carefully transitions from the GUT scale.  

On the other hand, the signal from the plasma does not require physics beyond the
SM and is thus also expected in the SM\@.
Apart from the already mentioned cosmic strings and other topological
defects, GW can also be produced in an incomplete phase transition \cite{Barir:2022kzo}, which could trigger
observable effects in the CMB\@.  We will briefly consider this option
in Section~\ref{sec:incompPT}.
Before turning to these various sources of GW, we will introduce the
concrete GUT scenario we consider, calculate the effective scalar
potential governing the FOPT, and discuss the viable parameter space in
Section~\ref{sec:Model}.

\section{SO(10) Model} \label{sec:Model}
\subsection{General Assumptions} \label{sec:Assumptions}
In order to shed light on the possible GW signals produced by GUTs in the early Universe, we adopt the following breaking chain pattern in this work:
\bea
\label{eq:bcsu3su2Rsu2Lu1BL}
SO(10)&\xrightarrow[45~[\mbox{\scriptsize{1}]} ]{\MG} &\left(SU(3)_C\times SU(2)_L \times SU(2)_R \times U(1)_{B-L} \right)/\mathbb{Z}_2
\nonumber\\
&\xrightarrow[126~[\mbox{\scriptsize{2}]} ]{\MR} & SU(3)_C \times SU(2)_L \times U(1)_Y  
  \xrightarrow[10]{\MEW} SU(3)_C \times U(1)_\text{em},
\eea
where we denote the appearance of monopoles and cosmic strings with [1] and [2],
respectively \cite{Lazarides:1980cc,Kibble:1982dd,Jeannerot:2003qv,Maji:2025yms}.
We choose this chain because it allows gauge coupling unification without introducing supersymmetry and is in compliance with the current experimental limits on the proton decay rate.
An additional appeal of this breaking chain is that, at least in its minimal version (with just the scalar content of~\eq{eq:bcsu3su2Rsu2Lu1BL}, plus three generations of fermions in the spinorial representation of $SO(10)$) it has been
shown to have promising phenomenological prospects~\cite{Bertolini:2009es,Bertolini:2012im,Graf:2016znk,Jarkovska:2021jvw,Jarkovska:2023zwv} (with the possible exception of some tension with electroweak observables that might be relaxed by extending the $\mathbf{10}$ representation at the last stage in \eq{eq:bcsu3su2Rsu2Lu1BL} with an extended Higgs sector).

\begin{figure}
    \centering
\includegraphics[width=0.7\linewidth]{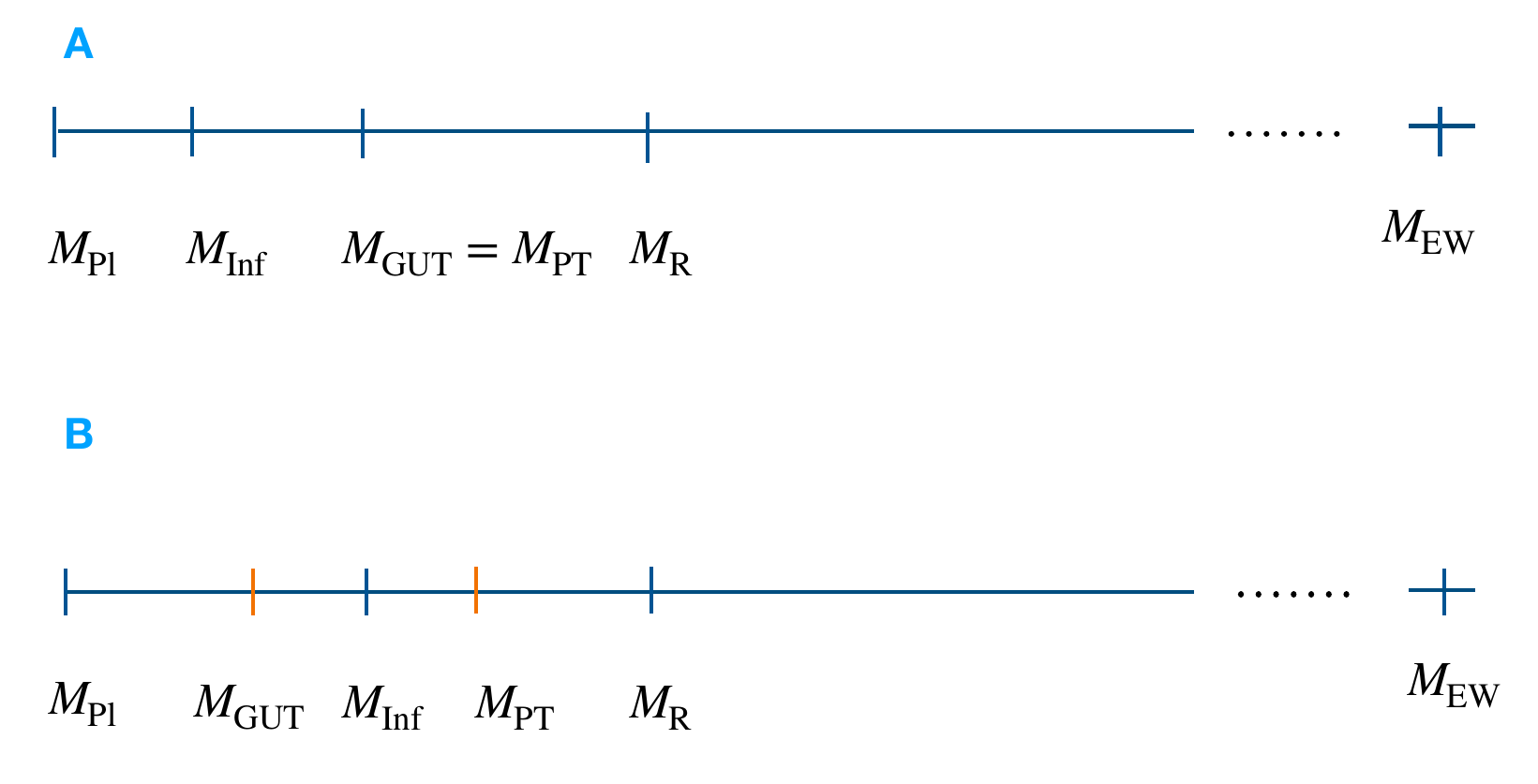}
    \caption{Possible scenarios for the completion of the phase transition. In scenario ``A'', the phase transition happens at the scale of grand unification and after the inflation scale, so its effects are visible today. Scenario ``B'' refers to the case that the phase transition does not complete at $\MG$ but is completed after inflation. We briefly mention this possibility in Section~\ref{sec:incompPT}. Here $\Mp$, $M_{\rm{Inf
}}$, $\MG$, $M_{\rm{PT}}$, $\MR$ and $\MEW$ are respectively the Planck mass, the inflation scale, the FOPT scale, the GUT scale, the scale at which $SU(2)_R$ is broken, and finally the EW scale.}
    \label{fig:possible_scale_order}
\end{figure}

In the following, we will focus on the breaking chain \eqref{eq:bcsu3su2Rsu2Lu1BL}, and the scenario ``A'' in \fig{fig:possible_scale_order} where the thermal evolution of the Universe after inflation starts in the $SO(10)$\footnote{Recall that formally we have Spin(10).}-symmetric phase.
The first step of symmetry breaking at the GUT scale $\MG$ leads to the production of monopoles.
In order to avoid overclosing the Universe, the monopole density has to be drastically reduced during the evolution below $\MG$.
A number of mechanisms are known that achieve this goal. The first one is embedding $SO(10)$ into a larger group and breaking that group to $SO(10)\times U(1)_\psi$ before inflation. If both $SO(10)$ and $U(1)_\psi$ are broken at the same scale, the monopoles appearing in the breaking of the former can attach to the strings from the breaking of the latter if the magnetic fluxes are configured appropriately, leading to their rapid annihilation \cite{Vilenkin:1982hm,Hindmarsh:1985xc, Preskill:1992ck,Blanco-Pillado:2007kxw,Kibble:2015twa,Lazarides:2019xai}. As a proof of principle, we give an explicit example realizing this possibility with an $E_7$ embedding in Appendix~\ref{app:flux}. Developing more elegant solutions would be an interesting direction for GUT model building, but we do not expect significant changes of the phenomenology of the scenario that we study here.
Alternatively, a period of thermal inflation could dilute the monopole density \cite{Lyth:1995ka,Lyth:1995hj}.
 In this case, the GW spectrum is also diluted and deformed \cite{Hu:2025xdt}, likely rendering the GW produced by the FOPT unobservable. However, the GW produced by the thermal plasma of the SM and the intermediate gauge group (see Section \ref{sec:Plasma}) will still be present. Hence, this case would require a separate analysis, which we do not pursue here.
Furthermore,
in some gauge-Higgs theories, by enlarging the scalar sector one may
admit embedded string solutions within the vacuum manifold.  These
strings can connect to or end at monopoles, so that the would-be
``free'' monopole is actually part of a monopole-string configuration.  Because the string can absorb part of the magnetic flux or change the boundary conditions, the isolated monopole loses its topological protection and can become unstable \cite{Vachaspati:1992fi,Achucarro:1999it}. The second step of symmetry breaking at $\MR$ of the particular chain of \eq{eq:bcsu3su2Rsu2Lu1BL} produces cosmic strings, which are topologically stable and produce GW that can be detectable \cite{Dunsky:2021tih,Chun:2021brv}.
We do not elaborate further on topological defects and their GW
signatures, which would be a source of GW production in addition to the mechanisms which are the main objective of our studies (i.e., the GW production by the FOPT and the thermal plasma).

It is worth pointing out that in this scenario the unification scale turns out to be close to the upper bound on the inflation scale 
$
V_\text{Inf} \sim (1.6 \times 10^{16}\GeV)^4
$
for slow-roll inflation driven by a single scalar with canonical
kinetic term \cite{Planck:2018jri}.
An upper bound of the same order of magnitude exists also for the temperature of the plasma after reheating, in order for the latter to be in thermal equilibrium~\cite{Kolb:1990vq}.
Given this tight space between inflation and the GUT scale, one can consider instead that unification takes place before inflation, as schematically depicted as option ``B'' in Fig.~\ref{fig:possible_scale_order}. In Section \ref{sec:potential} we describe the potential that would give rise to the FOPT and in Section~\ref{sec:incompPT} we discuss realizations of the option ``B'' in the context of transitions that start before inflation and end after inflation \cite{Barir:2022kzo}, leaving an imprint in the CMB and with effects in a broad bandwidth. 

\subsection{Definition and Matter Content}
We consider the minimal $SO(10)$ model containing only the gauge bosons,
the three families of fermions transforming under the representation
$\mathbf{16}$, as well as scalars transforming under $\mathbf{10}$,
$\mathbf{45}$, and $\mathbf{126}$
\cite{Graf:2016znk,Jarkovska:2021jvw,Jarkovska:2023zwv}.%
\footnote{In \cite{Jarkovska:2023zwv} it has been argued that is
challenging to accommodate all the necessary phenomenology for a successful
model, so in Sections \ref{sec:GCR} and \ref{sbsec:ProtonDecay} we
comment on possible departures from the minimum contents.}
We will refer to this model as $\mathbf{G}^{\rm M}_{3221}$, since the
first step of symmetry breaking leads to the gauge group
\begin{equation} \label{eq:groupsu3su2Lsu2RBL}
G_{3221}:=SU(3)_C \times SU(2)_L\times SU(2)_R \times U(1)_{B-L} \,.
\end{equation}
We will investigate the potential for a first-order phase transition in
this step, triggered by a vev $v$ of the $G_{3221}$-singlet component of
the $\mathbf{45}$ (see App.~\ref{sec:so10} for details).

As indicated in \eq{eq:bcsu3su2Rsu2Lu1BL}, we assume that a
$\mathbf{126}$ vev is solely responsible for the later
symmetry breaking to the SM gauge group, which implies that the
$SU(2)_R$ triplet component of the $\mathbf{45}$ does not obtain a vev,
i.e., $\omega_R=0$ in the notation of
\cite{Jarkovska:2021jvw,Jarkovska:2023zwv}.
We also assume that only the component of \rep{126} obtaining the vev
has a mass of $\mathcal{O}(M_R)$, while the masses of the other components 
are close to the unification scale \cite{Mambrini:2015vna,Biswas:2022cyh}.  The masses of the \rep{45} scalars
are of $\mathcal{O}(M_\text{GUT})$ as well.
For the multiplet \rep{10}, a large mass splitting is required
between the doublet and triplet components to avoid rapid proton decay,
cf.~Section~\ref{sbsec:ProtonDecay}.
Below $M_R$ we have exactly the gauge group and particle
content of the SM.

\subsection{Effective Potential \label{sec:potential}}
In this section, we present the main results for the contributions to the
effective scalar potential, including $1$-loop and thermal corrections.
Details of the computation are given in App.~\ref{app:Eff_Potential}.
We start with the tree-level $SO(10)$-symmetric
potential for the scalar $\mathbf{45}$,
\bea
\label{eq:V0tree45}
V_0(\phi) =
-\frac{\mu^2}{2} \Tr\phi^2 + \frac{a_0}{4} \left(\Tr\phi^2\right)^2 + \frac{a_2}{4} \Tr\phi^4
\eea
with $\mu^2 > 0$.
In terms of the classical field $\phi_c$, the tree-level potential
corresponding to the first step of the chain in \eq{eq:bcsu3su2Rsu2Lu1BL}, is
obtained by plugging \eq{eq:DefPhiC} into \eq{eq:V0tree45}, resulting in
\begin{equation} \label{eq:V0treephic}
V_0(\phi_c) =
-\frac{1}{2} \mu^2 \varphi_c^2 + a_0 \varphi_c^4 + \frac{1}{6} a_2 \varphi_c^4 \,.
\end{equation}
Defining the unification scale $v \equiv M_\text{GUT}$ as the
value of the classical field $\varphi_c$ at the potential minimum,%
\footnote{In the notation of \cite{Jarkovska:2021jvw,Jarkovska:2023zwv},
 this implies $v=\sqrt{3}\,\omega_{BL}$. Note that this vev is not at
 the global minimum of the tree-level potential, but this is possible
 once 1-loop corrections are taken into account \cite{Bertolini:2009es}.}
the tree-level minimization condition yields
\begin{equation} \label{eq:mu2a0a2tree}
	\mu^2 = \Bigl( 4 a_0+\frac{2}{3}a_2 \Bigr) \, v^2 \,.
\end{equation}

The $1$-loop contribution to the effective potential at zero temperature,
$V_1(\phi_c)$, contains two parts, the gauge boson contribution
\begin{equation} \label{eq:1loopgauge}
	V_1^g(\phi_c) = \frac{3}{64\pi^2} \sum_{i=g1,g2}
	n_i \, m_i^4(\phi_c)
	\left(\ln\frac{m_i^2(\phi_c)}{\mu_r^2}-\frac{5}{6}\right) ,
\end{equation}
and the scalar contribution
\begin{equation} \label{eq:1loopscalar}
	V_1^s(\phi_c) = \frac{1}{64\pi^2} \sum_{i=s1,s2,s3,\chi}
	n_i \, m_i^4(\phi_c)
	\left(\ln\frac{|m_i^2(\phi_c)|}{\mu_r^2}-\frac{3}{2}\right) ,
\end{equation}
where $m_i(\phi_c)$ is the field-dependent mass of the gauge boson or scalar $i$ and
$n_i$ the number of particles with this mass.  Both quantities are given
in \tab{tab:FieldDepMasses}.
These expressions have been obtained in the $\overline{\text{MS}}$
renormalization scheme with the renormalization scale $\mu_r$, for which
we choose $\mu_r=v$ in numerical calculations.
Note that the scalar contribution contains both the physical scalars and the
Nambu-Goldstone bosons.
We neglect the contributions of the scalars from representations other
than \rep{45}, since they would introduce many additional parameters but
are not expected to have a sizable impact on the phase transition.
Note that the scalar potential that we are considering here for the
first step of the symmetry breaking is independent of any physics below
the unification scale, of course apart from the value of the unification
scale, which is determined by the breaking chain and the matter content.

\begin{table}
\centering
\begin{tabular}{cccc}
\multicolumn{2}{c}{Field-dependent mass squared} & Particle type & Multiplicity $n_i$ \\
\hline
$m_{g1}^2(\phi_c)$ & $\frac{1}{6} g^2 \varphi_c^2$ & Gauge boson & $24$ \\
$m_{g2}^2(\phi_c)$ & $\frac{2}{3} g^2 \varphi_c^2$ & Gauge boson & $6$ \\
$m_{s1}^2(\phi_c)$ & $-\mu^2 + 4 a_0 \varphi_c^2$ & Scalar & $6$ \\
$m_{s2}^2(\phi_c)$ & $-\mu^2 + 12 a_0 \varphi_c^2 + 2 a_2 \varphi_c^2$ & Scalar & $1$ \\
$m_{s3}^2(\phi_c)$ & $-\mu^2 + 4 a_0 \varphi_c^2 + 2 a_2 \varphi_c^2$ & Scalar & $8$ \\
$m_\chi^2(\phi_c)$ & $-\mu^2 + 4 a_0 \varphi_c^2 + \frac{2}{3} a_2 \varphi_c^2$ & NGB & $30$ \\
\hline
\end{tabular}
\caption{Field-dependent masses.  The multiplicity indicates the number
of particles with this mass. The gauge coupling $g$ is the gauge coupling of the SO(10) theory, defined in \eq{eq:fieldstrength}.}
\label{tab:FieldDepMasses}
\end{table}

In \eq{eq:Vthermal} we specify the $1$-loop contributions to the
effective potential at finite temperature, $V_{\rm{th}}(\phi_c,T)$.
Then the complete $1$-loop effective potential is given by
\begin{equation}
	V(\phi_c,T) =
	V_0(\phi_c) + V_1^g(\phi_c) + V_1^s(\phi_c) + V_\text{th}(\phi_c,T) \,.
\label{eq:completepot}
\end{equation}

We need to find the minimum
of the potential including these corrections.
The condition for the minimum at zero temperature is
\bea
\label{eq:VSO10eff}
\frac{\partial V(\phi_c,0)}{\partial \phi_c}=\frac{\partial V_0(\phi_c)}{\partial \phi_c}+ \frac{\partial V_1^g(\phi_c)}{\partial \phi_c}+ \frac{\partial V_1^s(\phi_c)}{\partial \phi_c}=0,
\eea
which can be solved numerically.
What we adopt here is a similar strategy to the one followed in
\cite{Jarkovska:2021jvw} in that we adjust the value of $\mu^2$ such that
the minimum at the $1$-loop level is roughly equal to $v$.
We remind the reader that $v$ is fixed by the gauge coupling unification
condition, as we will discuss in Section~\ref{sec:GCR}.
Specifically, we calculate $\mu^2$ at the tree level from
\eq{eq:mu2a0a2tree} and use it in \eq{eq:VSO10eff} to find the value of
the scalar field in the minimum, denoted by $\phi_m$.  This value
differs from $v$, so we update $\mu^2$ iteratively using Newton's method
until $\phi_m$ differs from $v$ by no more than $1\,\%$.
As the value of $\mu^2$ thus obtained approaches zero, the calculation becomes numerically challenging due to convergence issues and large higher-order loop corrections.
Therefore, we exclude points with $\mu^2<0$.

Given the size of the gauge coupling at the GUT scale (around $0.5$ as we explain
in Section~\ref{sec:GCR}) and the particle content of the model (thirty
massive gauge bosons and fifteen scalar bosons), we expect that loop
corrections can become big and one-loop precision may not be
sufficient.
As a two-loop computation is beyond the scope of this work, we adopt the
usual strategy for estimating the uncertainty caused by neglecting
higher loop orders, see e.g.~\cite{Buras:1998raa,Plehn:2009nd},
defining the quantity 
\begin{equation} \label{eq:Delta2Loop}
\Delta^\text{2-loop} := \left\langle
\frac{\left.V(\phi_c,0)\right|_{\mu_r=2v}-\left.V(\phi_c,0)\right|_{\mu_r=v/2}}{\big(\left.V(\phi_c,0)\right|_{\mu_r=2v}+\left.V(\phi_c,0)\right|_{\mu_r=v/2}\big)/2}
\right\rangle_{\!0<\varphi_c<1.2v}
\end{equation}
to estimate the importance of higher loop orders, since the exact potential is
independent of $\mu_r$.
The brackets in \eq{eq:Delta2Loop} indicate an average over $1000$
linearly sampled values of $\varphi_c$ between 0 and $1.2\,v$ to reduce the sensitivity to single points in field space where the denominator becomes very small due to cancellations.%
\footnote{Such cancellations can occur if a potential barrier is present in $V(\phi_c,0)$, since then the potential necessarily crosses zero between the minimum at the origin and the global minimum.}

Before presenting the parameter space where a first-order phase
transition would be possible, we discuss in the next sections
constraints on the model that lead to bounds on the unification scale
and on the parameter space of the coefficients in the potential.

\subsection{Gauge Coupling Running \label{sec:GCR}}

The minimal version of the model with the breaking chain of \eq{eq:bcsu3su2Rsu2Lu1BL} contains the scalar $\mathbf{45}$ which breaks the $SO(10)$ group, then the ${\mathbf{126}}$ breaks subsequently to $G_{SM}$ and then the $\mathbf{10}$ causes the breaking at the electroweak (EW) scale. Our convention for the order of the indices is 
 $a=3, 2L, 2R, B-L$  for this model the component acquiring a vev in the $\mathbf{126}$ is $(1,1,3,2)$ and that of $\mathbf{10}$ is $(1,2,2,0)$ $\supset (2,1,\pm 1/2)$, where this last decomposition is  under the SM group, $SU(2)_L\times SU(3)_C \times U(1)_Y$.
For the matter, we have the three families of fermions: $(3,2,1,-1/3)$, $(1,2,1,1)$, $(\overline{3},1,2,1/3)$ and $(1,1,2,-1)$. However, we just take into account the running of the top mass up to the $\MR$ scale. With this minimal matter content, the beta function coefficients at one and two loops are, respectively \cite{Mambrini:2015vna,Chakrabortty:2019fov,Chun:2021brv},
\bea
\label{eq:bfuncURUBLNoDNS}
b_a^{(1)}=\left(
\begin{array}{c}
-7\\
-3\\
-7/3\\
11/2\\
\end{array}
\right),\quad
b_{ab}^{(2)}=\left(
\begin{array}{cccc}
-26  & 9/2 & 9/2 & 1/2\\  
12 & 8 & 3 & 3/2 \\
12  & 3 & 80/3 & 27/2 \\
12 & 9/2 & 81/3 & 61/2\\
\end{array}
\right).
\eea
The matching conditions at the scale $\MR$ are
\bea
\label{eq:matchingMR}
g^{SM}_{2}(\MR) &=& g^{G_{3221}}_{2L}(\MR) \,, \nonumber\\
g^{SM}_1(\MR)&=&\left[\frac{3}{5} \frac{1}{{g^{G_{3221}}_{2R}}^2(\MR)} + \frac{2}{5}\frac{1}{{g^{G_{3221}}_{B-L}}^2(\MR)}\right]^{-1/2}, \nonumber\\
g^{SM}_{3}(\MR) &=& g^{G_{3221}}_{3}(\MR) \,,
\eea
where $g^{G_{3221}}_{2L}$, $g^{G_{3221}}_{2R}$
$g^{G_{3221}}_{B-L}$ and $g^{G_{3221}}_{3}$ are respectively the couplings of the group factors $SU(2)_L$, $SU(2)_R$, $U(1)_{B-L}$ and $SU(3)_C$.
We use the standard GUT normalization for the $U(1)$ gauge couplings.
To unclutter the notation, from now on we suppress the superscript $G_{3221}$.

When running up the SM gauge coupling and matching to the $G_{3221}$ group we need to make a guess about the relation between $g_{2R}$ and $g_{2L}$ but after running back and forth, in the final run from $\MG$ to $\MR$, we can then obtain the final predicted value for $g_{2R}(\MR)$ and $g_{2L}(\MR)$.
The beta functions take the form 
$dg_a/dt =g_a^3 b_a/16\pi^2+g_a^3/(16\pi^2)^2\left[\sum_{b=1}^4 b_{ab} g_b^2  \right]$, for $a=3, 2L, 2R, B-L$ and the coefficients $b_a$ and $b_{ab}$ as in \eq{eq:bfuncURUBLNoDNS}. For the SM,
$dg_a/dt =g_a^3 b_a/16\pi^2+g_a^3/(16\pi^2)^2\left[\sum_{b=1}^3 b_{ab} g_b^2 - C_a^t y_t^2 \right]$ with $C_1=17/10$, $C_2=3/2$, $C_3=2$, for $a=1,2,3$ representing respectively the SM group factors $U(1)_Y, SU(2)_L, SU(3)_C$. The coefficients $b_a$ and $b_{ab}$ for this case are very well known and can be found in \cite{Machacek:1984zw}.
Just with this minimal content the scales are
\bea
\label{eq:45_NONSUSY_Scales}
\MR^{2\, \rm{loop}} &=& (2.9\pm 1.0) \times 10^{9}\GeV \,,
\nonumber\\
\MG^{2\, \rm{loop}} &=& (1.6\pm 1.0) \times 10^{16}\GeV \,,
\eea
where the uncertainties stem mainly from threshold corrections and numerical precision. Once more constraints are taken into account \cite{Jarkovska:2023zwv}, the scale $\MG$ can be as low as $\mathcal{O}(10^{15})\GeV$. However, finding all phenomenology to be viable, especially in the Higgs sector at the EW scale \cite{Jarkovska:2023zwv}, seems challenging. Adding matter content impacts the running and so the unification scale. Therefore, in what we consider here we set values of the unification scale in the range from $10^{15}\GeV$ to $1.6 \times 10^{16}\GeV$. 
\begin{figure}
    \centering
\includegraphics[width=0.49\linewidth]{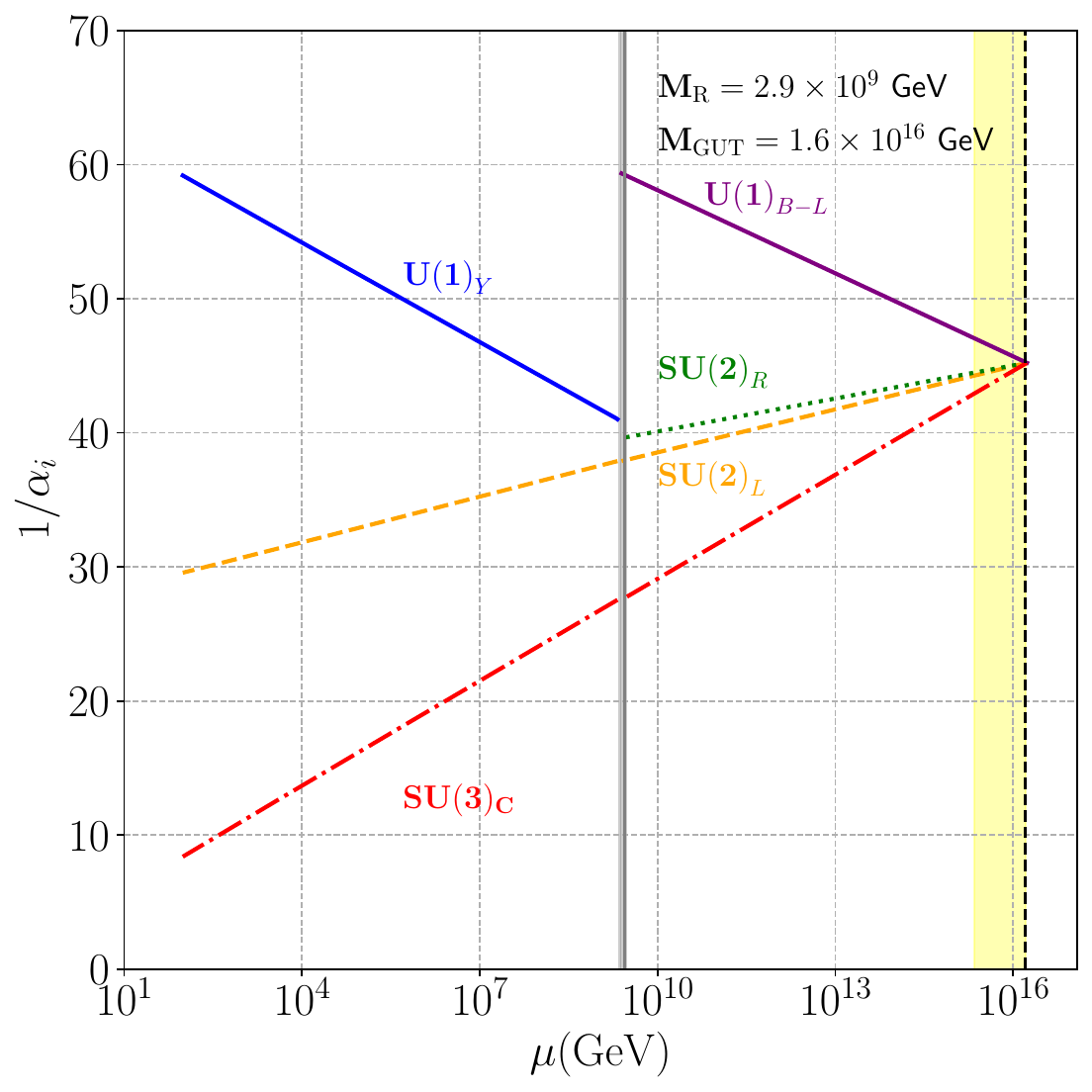}
\includegraphics[width=0.49\linewidth]{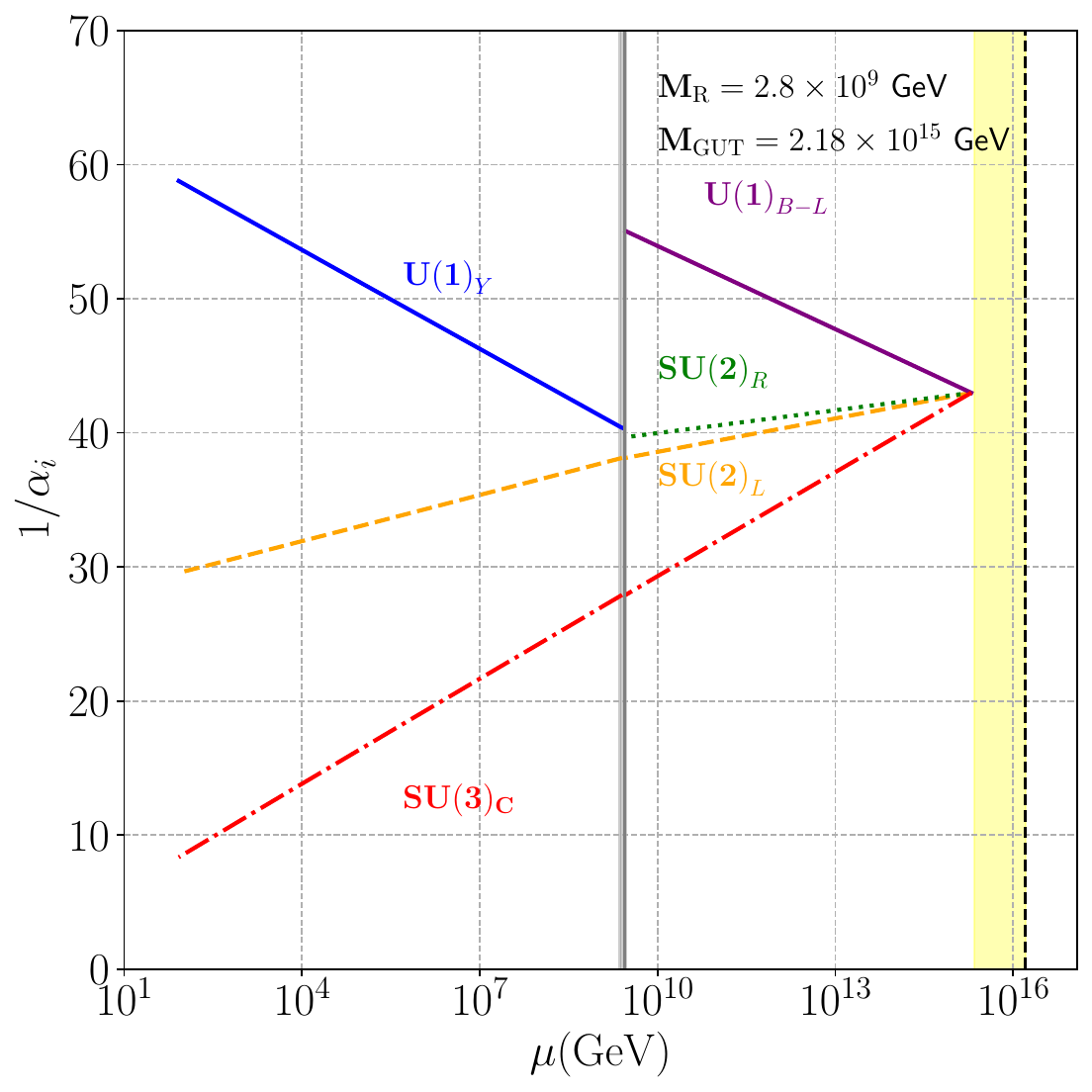}
    \caption{The left panel represents the running of the  Minimal
$\mathbf{G}_{3221}$ model, $\mathbf{G}^{\rm M}_{3221}$ (containing only
$\mathbf{45}\oplus\mathbf{126}\oplus \mathbf{10}$), with  $\MR=(2.9\pm
1.0) \times 10^{9}\GeV$ and $\MG= (1.60\pm 1.0) \times 10^{16}\GeV$. The
gauge couplings of the groups of the SM are run up to $\MR$ and those of
$SU(3)_C\times SU(2)_L \times SU(2)_R\times U(1)_{B-L}$ from $\MR$ and
$\MG$. The vertical gray band around $2.8\times 10^9\GeV$ represents the band allowed for $\MR$ for this model and the model of the right plot.  The right plot represents the running of the 
$\mathbf{G}_{3221}^\text{M}$ model plus a fermionic $SU(2)_R$-$SU(2)_L$ bi-doublet Dark Matter candidate (that of the third panel Fig.~6 of \cite{Biswas:2022cyh}). We present this last plot as an example that the addition of matter can alter the running  (see \eq{eq:betagigen} and the text below) of the gauge coupling constants and reduce the unification scale.}
    \label{fig:ex_running}
\end{figure}
Since it is challenging to accommodate the EW physics in the minimal $G^{\rm M}_{3221}$, we may consider the addition of more scalars or fermions. For this purpose, we recall the general expression for the 1-loop beta function
of the gauge coupling of the group $\mathcal{G}_n$,
\begin{equation} \label{eq:betagigen}
\beta_{g_n}=\frac{g_n^3}{16\pi^2} b_n \quad,\quad
b_n = -\frac{11}{3} I(\text{ad}^{(n)}) + \frac{2}{3} \sum_i I(F_i^{(n)}) + \frac{1}{6} \sum_j I(S_i^{(n)})
\,,
\end{equation}
where $\text{ad}^{(n)}$, $F_i^{(n)}$, and $S_j^{(n)}$ denote the representations of
$\mathcal{G}_n$ transforming the gauge bosons, chiral fermions, and real
scalars, respectively, and $I$ is the Dynkin index of a representation.
As is well known, for the SM $SU(3)_C$ factor, we have
$I(\text{ad}^{(3)})=3$ and $\sum_i I(F_i^{(3)})=6$, rendering $b_3=-7$.
We can see that the addition of particles that make $b_3$ less negative, for example, colour triplets, flattens the running of $1/\alpha_3$ and shifts $\MG$ to higher scales and $\MR$ to lower scales. But as some of us noted in \cite{Biswas:2022cyh}, the addition of colour triplets is not enough because multiplets of all the other groups are required in order to ensure that the couplings 
$g_{2L}$ and $g_{2R}$ unify at high scale.
As an example, in \fig{fig:ex_running} we plot the running of the gauge
couplings when a Dark Matter candidate in the form of a self-conjugate
fermion with $B-L=0$ which is a doublet of both $SU(2)_R$ and $SU(2)_L$
(referred to as ``bi-doublet'') and a singlet of $SU(3)_C$ has been
added to the model $G_{3221}^\text{M}$.
For this case the GUT scale is located at $2.8\times 10^{15}\GeV$. Further examples of adding Dark Matter candidates can be found in \cite{Mambrini:2015vna,Biswas:2022cyh}. Note that in our analysis of FOPT, we consider both the maximum scale $\MG = 1.6\times 10^{16}\GeV$ and the scale $\MG=10^{15}\GeV$ because that scale is still compatible with proton decay. One further justification for this is that we can add matter at lower energies, that will alter the scale of gauge coupling unification but not the basic characteristics of the potential of \eq{eq:V0tree45}.

\subsection{Constraints within the Model}
\subsubsection{Mass Spectrum}
Calculating the mass spectrum arising from the GUT-scale symmetry
breaking step, one encounters particles with negative mass-squared, i.e.,
tachyons.  However, it is necessary to
take into account loop corrections to the masses.
 At the one-loop level, this entails adding all terms involving the \rep{126} scalar to the potential in \eq{eq:V0tree45}, determining the effective potential by extending the sums in Eqs.~(\ref{eq:1loopgauge},\ref{eq:1loopscalar}) to the full particle content, and calculating the eigenvalues of the scalar mass matrix, which is only possible numerically. This was done in \cite{Jarkovska:2021jvw}, where it was shown that the tachyon problem can be avoided if the quartic couplings of the scalar \rep{45} are restricted to
 (see Figs.~4 and 6, Section~IV.C.1 point (iii), and Appendix~A of \cite{Jarkovska:2021jvw})
\begin{equation} \label{eq:a0a2range}
	a_0 \in ( 0.0, 0.2) \quad,\quad
	a_2 \in (-0.05,-0.01) \,.
\end{equation}
Hence, we will only consider parameter values within these ranges in our
analysis.
This also ensures that the vev of the \rep{45} is
located at the global minimum of the 1-loop effective potential.

\subsubsection{Proton Decay \label{sbsec:ProtonDecay}}
Here we follow \cite{Mambrini:2015vna,Ellis:2019fwf,Chun:2021brv} and give a brief account on the way proton decay rates are calculated.  The most sensitive channel for non-superysmmetric models is the dimension-6 operator induced decay $p\rightarrow \pi^0  \, e^+$ which is mediated by gauge fields\footnote{Colour triplet scalars can also induce proton decay but since the dominant decay in non-supersymmetric models are gauge interactions we do not consider them here.}. The proton decay for this channel can be estimated as \cite{Mambrini:2015vna}
\bea
\label{eq:pdePi0}
\Gamma(p\rightarrow \pi^0  \, e^+)&=&
\frac{m_p}{32\pi}\left(1-\frac{m^2_{\pi^0}}{m^2_p} \right)^2\, 
\left[ |\mathcal{A}_L(p\rightarrow \, \pi^0 \, e^+ )|^2 + |\mathcal{A}_R(p\rightarrow  \, \pi^0 \, e^+)|^2
\right],
\eea
where $m_p$ and $m_{\pi^0}$ are respectively the proton and the neutral pion masses.
The amplitudes at the weak scale are given by 
\bea
\label{eq:Amplitudes_NS_PD}
{\mathcal{A}_L}(p\rightarrow \pi^0  \, e^+) & =& C_{RL}((ud)_R u_L)(\mu=2\, \rm GeV) 
\langle {\it{ 
\pi}}^0{\it{| (ud)_R u_R|p}}
\rangle, \nonumber\\
{\mathcal{A}_R}(p\rightarrow \pi^0  \, e^+)  &=& 2 C_{LR}((ud)_L u_R)(\mu=2\, \rm GeV) 
\langle {\it{
\pi}}^0{\it{| (ud)_R u_R|p}}
\rangle,
\eea
where the Wilson coefficients $C_{RL}((ud)_R u_L)$ and $C_{LR}((ud)_L u_R)$ correspond respectively to $C_1$ and $C_2$ of
\cite{Mambrini:2015vna}. For numerical values we use the inputs given in Tab.~1 in \cite{Chun:2021brv}, including the value of the hadronic matrix elements and the evolution of the Wilson coefficients from $\MG$ is given in \cite{Ellis:2019fwf}. For the minimal $G_{3221}$ model, we obtain $(2.4 \pm 1.6)\times 10^{36}$ years, a safe value given the current experimental bound of $\tau(p\rightarrow \pi^0 e^+) > 2.34 \times 10^{34}$ years at 95\% C.L.~\cite{Super-Kamiokande:2020wjk}
and the projected bound of $7.8 \times 10^{34}$ years at 95\% C.L.~\cite{Hyper-Kamiokande:2018ofw}.

\section{Appearance of a First Order Phase Transition} \label{sec:FOPT}
\subsection{Preliminaries}

In this section, we briefly mention only the salient concepts pertaining FOPT, which have been studied extensively (see for example \cite{Hindmarsh:2020hop}), in order to explain our results.  As it is well known, FOPT can generate GW through nucleation, expansion, collision, and merger of bubbles in the broken phase. In vacuum, GW originate solely from bubble collisions and can be described by the envelope approximation \cite{Huber:2008hg}. In a thermal plasma, however, friction slows bubble expansion, transferring most of the energy to the plasma and making scalar field contributions subdominant.
Lattice simulations show that this energy drives sound waves in the plasma, leading to an acoustic phase that dominates GW production \cite{Hindmarsh:2013xza,Hindmarsh:2015qta}. For stronger transitions, turbulence may emerge, sustaining GW emission until it dissipates \cite{Kosowsky:2001xp,Nicolis:2003tg,Caprini:2006jb,PhysRevD.86.103005,Kisslinger:2015hua}.
To model these effects, we use GW templates from 3D simulations that fit the spectrum in terms of phase transition parameters and frequency \cite{Caprini:2009yp,Binetruy:2012ze,Ellis:2018mja}. Key quantities characterizing the GW are $\alpha$ and $\beta$, respectively, the ratio of released vacuum energy to the plasma's radiation energy in the symmetric phase and a measure of the duration of the phase transition. For completeness, the formulas that we use for $\alpha$ and $\beta$ are presented in \eq{eq:defalpha} and \eq{eq:defbeta}, while the 3D action, $S_3$, describing the bubbles forming when the transition between the meta-stable and the true vacuum takes place is given in \eq{eq:S3T}. According to \cite{Coleman:1977py,Linde:1981zj,LINDE198137} the decay rate, $\Gamma(T)$, of a bubble is given by 
%
%
%
\bea
\label{eq:gammaT}
\Gamma(T) \approx T^4\left(\frac{S_3}{2 \pi T}\right)^{\frac{3}{2}} \exp\left(-S_3 / T\right).
\eea
Taking the first expression to be valid we can calculate the nucleation temperature $T_n$,   at which the average number of bubbles nucleated per Hubble horizon is of order 1:
\begin{align}
    N(T_n) = \left( \frac{3\Mpbar}{\pi}\right)^4\left( \frac{10}{g_{*}}\right)^2\int_{T_n}^{T_c} \frac{d T}{T^5} \left( \frac{S_3}{2 \pi T}\right)^{3/2} \exp{(-S_3/T)} \sim 1 \,,
\label{eq:NTnO1}    
\end{align}
which is equivalent to computing the nucleation temperature by taking
$\Gamma/V$, where $V$ is the comoving volume, as 
\begin{equation}
\frac{\Gamma}{V}\approx T_n^4 e^{S_3(T_n)/T_n} \,,
\label{eq:gammaoverv}
\end{equation}
integrating to the time $t_*$ at which the phase transition takes place and requiring that the total number of bubbles nucleated from time $t=0$ to $t=t_*$ is of order 1.\footnote{$\int_0^{t_*} dt \Gamma/v/H^3(t)=\mathcal{O}(1)$.}
\eq{eq:gammaoverv} has no solution when 
\begin{equation}
\label{eq:GammaVLH4}
\frac{\Gamma}{V}< H^4 \,,
\end{equation}
which can be written as
\bea
\label{eq:condition_inc_FOPT}
\frac{S_3(T)}{T}>4\log\left(\frac{\Mpbar}{T}\right)-2 \log\left(\frac{90}{\pi^2 g_*(T)}\right),
\eea
where $\Mpbar=\Mp/\sqrt{8\pi}=2.4\times 10^{18}\GeV$ is the reduced Planck mass.
In this case the transition is not completed.
For finding the suitable parameter space where a FOPT may occur, we
consider the one-loop corrected potential with thermal corrections in
\eq{eq:completepot} and the criterion \eq{eq:NTnO1} to determine whether nucleation is possible.
We note that for the potential that we are considering, the temperature
correction always grows and has a different sign with respect to the
tree-level and 1-loop contributions, which need to have a minimum and
become negative. In this regard, the thermal contribution is typically necessary to produce a barrier.  As is well understood, a barrier can occur if there is a cubic term in the field in the 1-loop thermally corrected potential (where the total contribution to the potential is $\phi^3 E$, where $E$ has mass dimension $1$ and could be temperature-independent). $E$ is proportional to gauge coupling of the theory to the third power, $g^3$, and hence the appearance of a barrier enhances with $g^3$. 

Above the unification scale, the considered model contains $96$
fermionic ($3$ generations in the representation $\mathbf{16}$, $2$
possible spin orientations) and $307$ scalar degrees of freedom ($45$
massless gauge bosons, $10 + 45 + 126 \times 2$ real scalars), which
yields $g_* = 481$ effective relativistic degrees of freedom.  However,
nucleation happens at temperatures below $M_\text{GUT}$, where some of
the particles obtaining GUT-scale masses are non-relativistic.  To
account for this, we use $g_*=300$ in our computations.

\subsection{Results \label{subsec:results}}
\begin{figure}
    \centering
   \hspace*{-0.45cm} \includegraphics[width=0.46\linewidth]{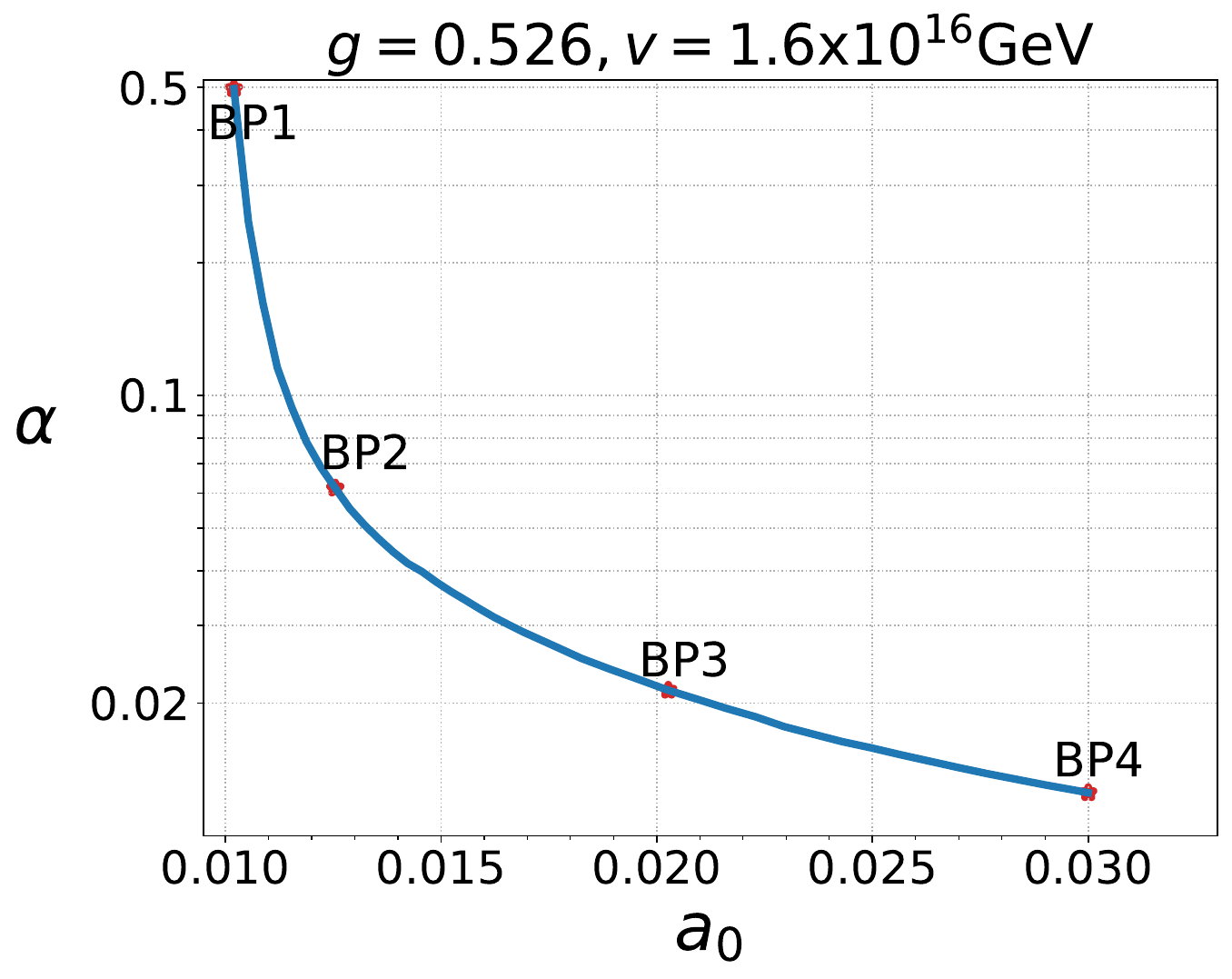} \hspace*{0.34cm}
 \includegraphics[width=0.46\linewidth]{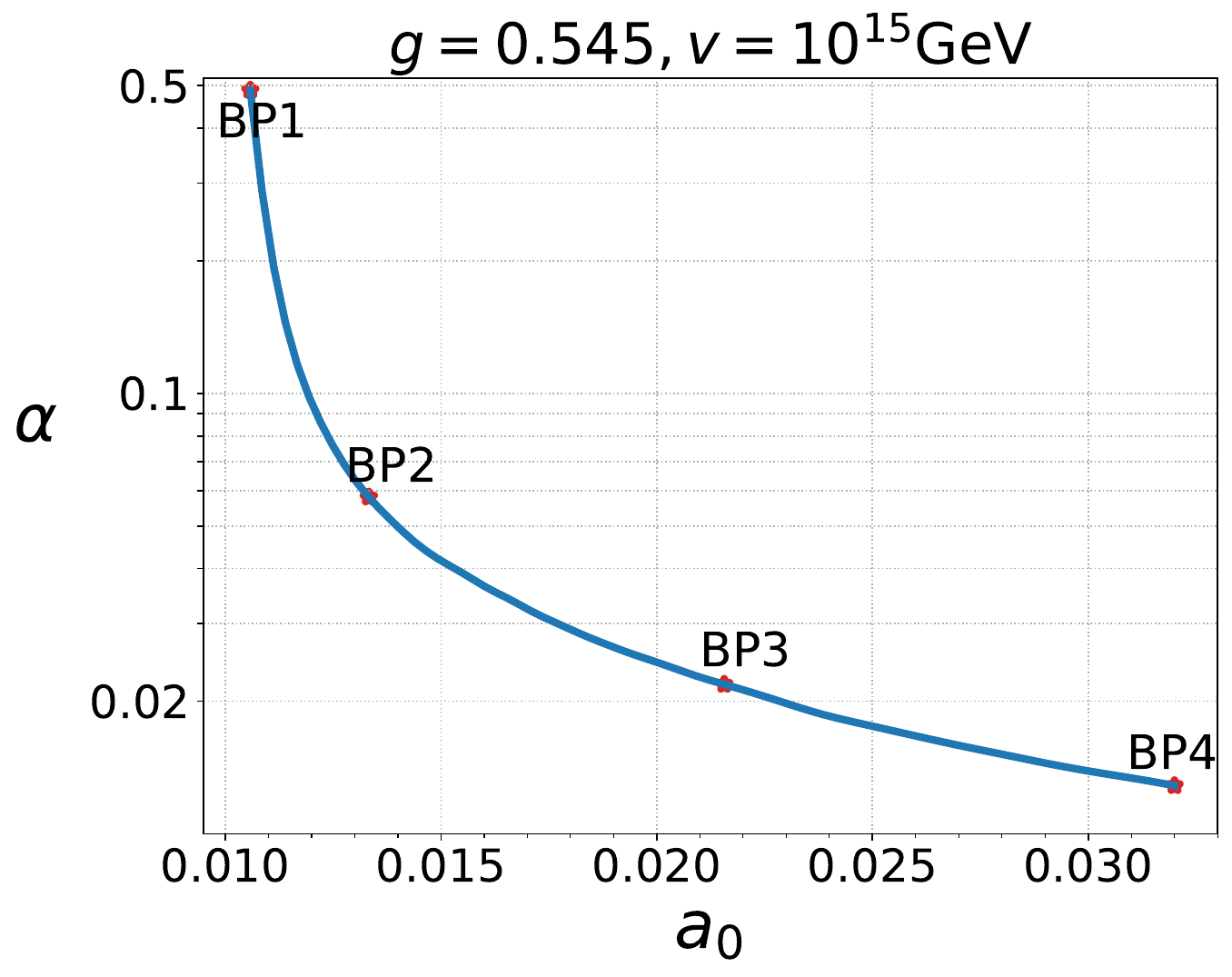}   
   \hspace*{-0.45cm}  \includegraphics[width=0.47\linewidth]{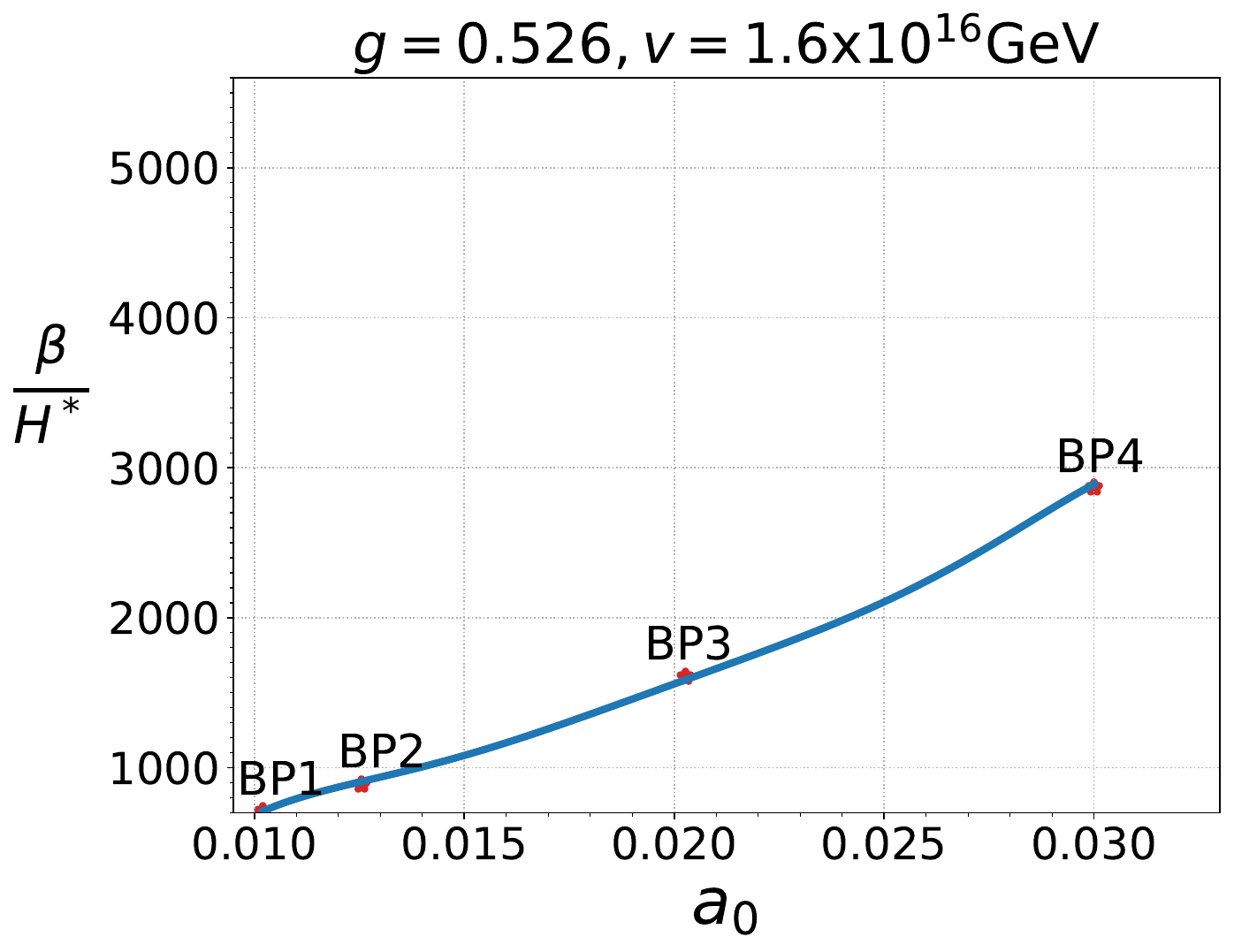} \hspace*{0.32cm}
    \includegraphics[width=0.47\linewidth]{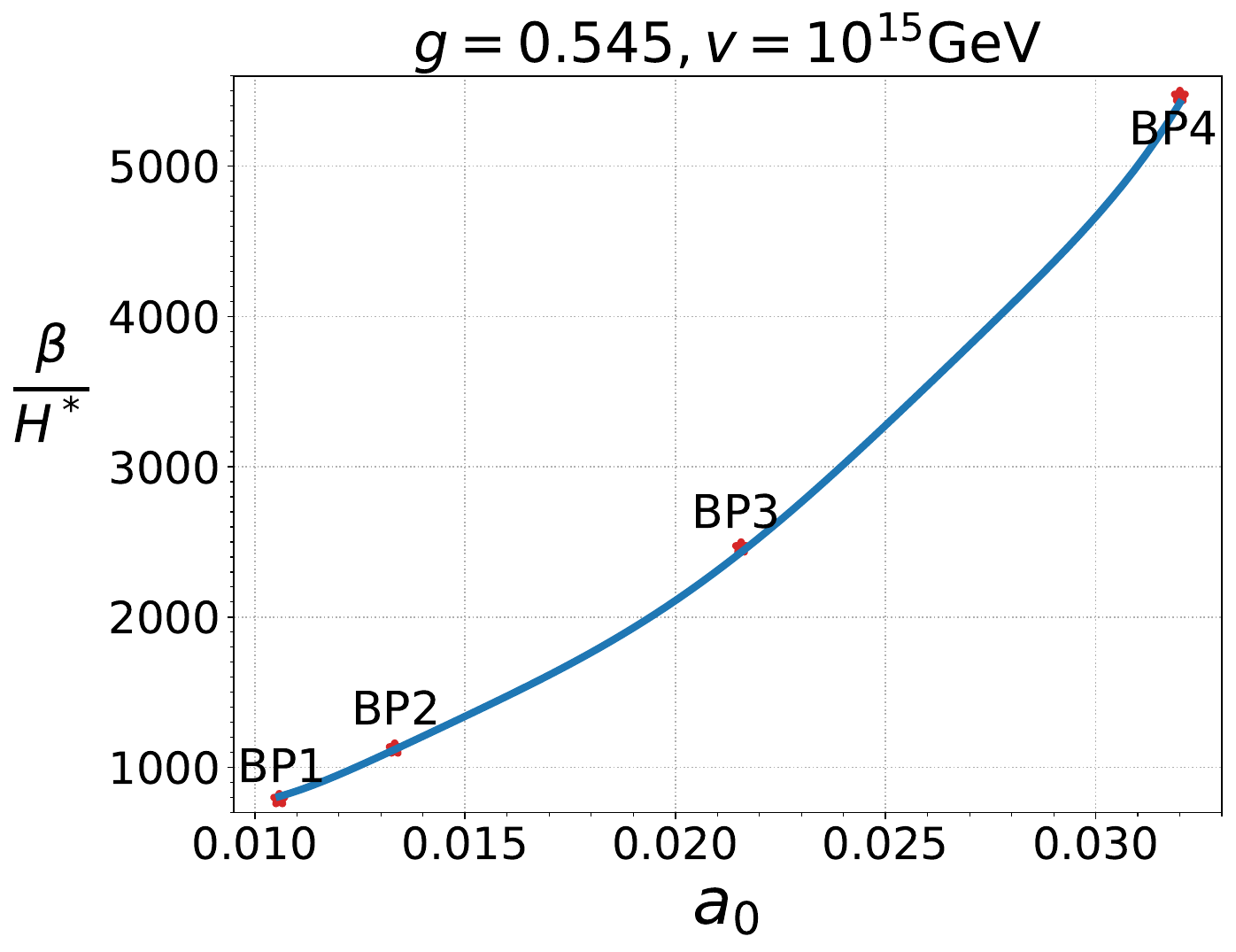}
    \hspace*{-0.9cm}
    \includegraphics[width=0.47\linewidth]{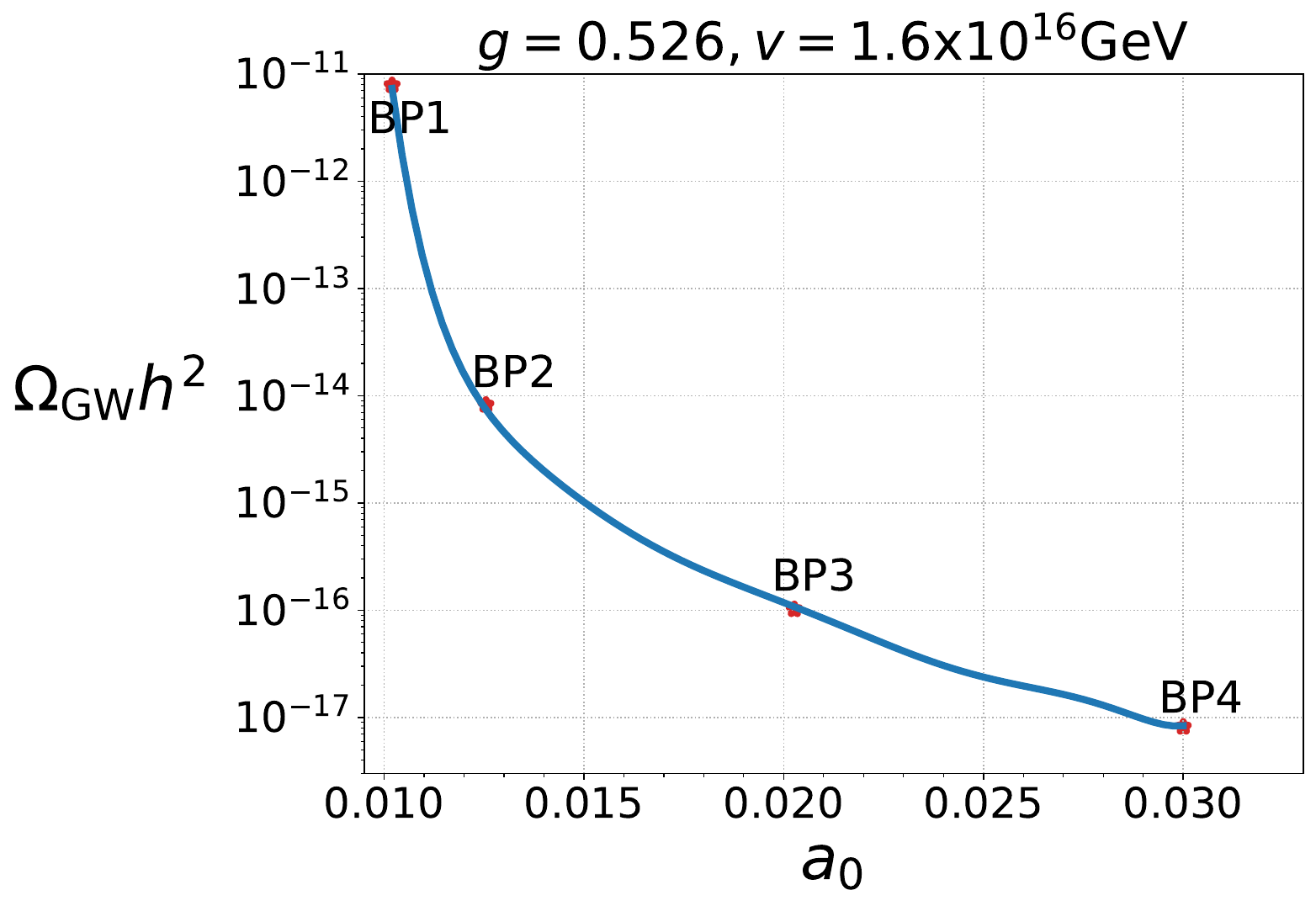}
    \includegraphics[width=0.47\linewidth]{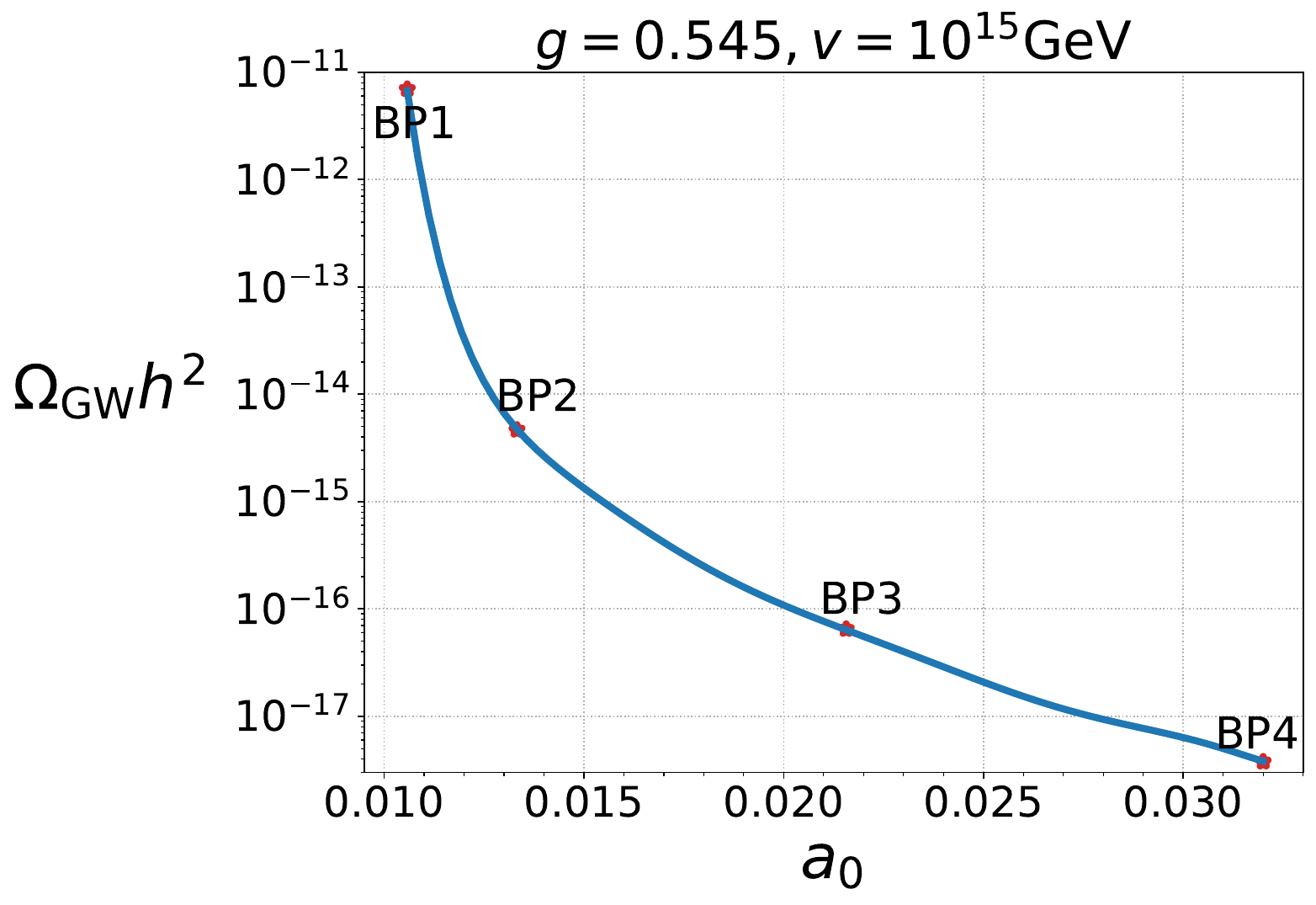}    
    \caption{Phase transition parameters ($\alpha$, $\beta/H_*$) and GW density as a function of $a_0$ for the scales $\MG=1.6 \times 10^{16}\,$GeV, left, and $\MG=1.0 \times 10^{15}\,$GeV, right. The (red) dots represent the benchmark points listed in Tables~\ref{tbl:BP_v1016GeV} and~\ref{tbl:BP_v1015GeV}.} 
\label{fig:PTparameters}
\end{figure}
We start by discussing the behaviour of the phase transition parameters
$\alpha$ and $\beta/H_*$, which indicate the strength and the inverse
duration, respectively, see Appendix~\ref{app:FOPT_conv}.
In \fig{fig:PTparameters} we collect the results for the scales $\MG=1.6\times 10^{16}\GeV$ and $\MG=1.0\times 10^{15}\GeV$. As is known, $\alpha$ and $\beta/H_*$ are not completely independent parameters, since both of them depend on the 3D action, \eq{eq:S3T} and they have opposite behaviours, that is, when $\alpha$ increases, $\beta/H_*$
decreases. This behaviour can be clearly seen in \fig{fig:PTparameters}, where we plot $\alpha$ and $\beta$ as a function of the parameter $a_0$, \eq{eq:V0treephic}, and whose range is constrained, such as to avoid a tachyonic spectrum, \eq{eq:a0a2range}. We have also marked four representative benchmark points (BP) and we give the critical, nucleation temperature, $\alpha$ and $\beta$ in \tab{tbl:BP_v1016GeV}. In addition, in \fig{fig:PTparameters}, we have plotted the GW energy density as computed with the equations of Appendix \ref{app:FOPT_conv}. Since $\Omega_\text{GW} h^2$ is proportional to $\alpha^2/(\beta/H_*)$, $\Omega_{GW}h^2$ follows a decreasing behaviour as $a_0$ grows, just as in the case of $\alpha(a_0)$. For each point in $a_0$, $\mu^2$ and $a_2$ are computed following the algorithm described in Section \ref{sec:potential}.

\begin{figure}
    \centering
    \includegraphics[width=0.8\linewidth]{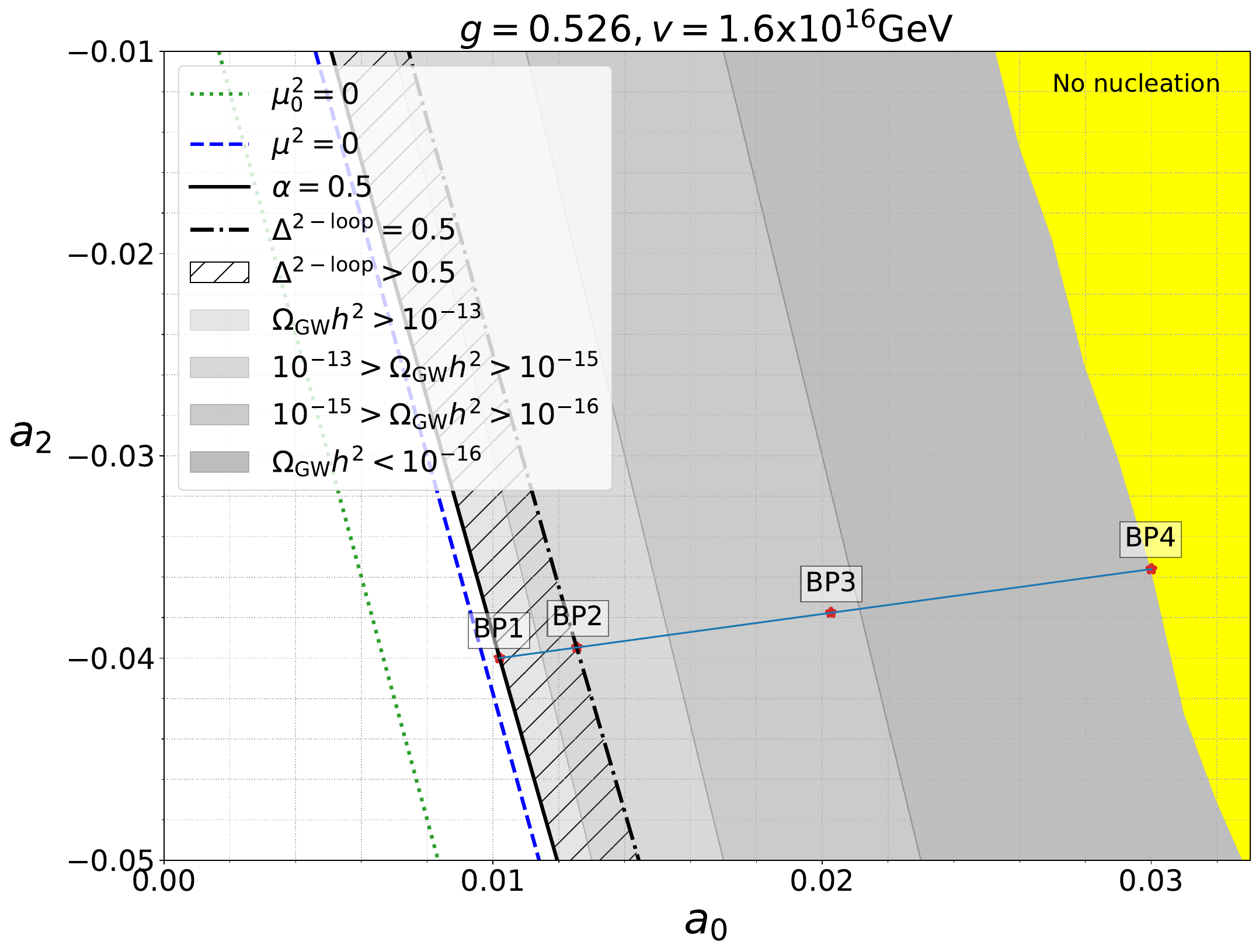}
    \caption{GW parameter space for $a_0$ vs.\
$a_2$, where $a_2$ is restricted to the region where the spectrum does
not contain tachyons. The four points marked by stars are our benchmark
points. The yellow region has no FOPT or is numerically difficult to
calculate. In the hatched region, $\Delta^\text{2-loop}>0.5$.
We denote the quadratic term in the scalar potential obtained from the
tree-level condition \eqref{eq:mu2a0a2tree} by $\mu_0^2$.  In contrast,
$\mu^2$ is the corrected value of this parameter ensuring that the
$1$-loop-corrected scalar potential has a minimum at $v$.}
\label{fig:cvgce_of_min1loop_1_1p6e16}
\end{figure}

\begin{figure}
    \centering
    \includegraphics[width=0.8\linewidth]{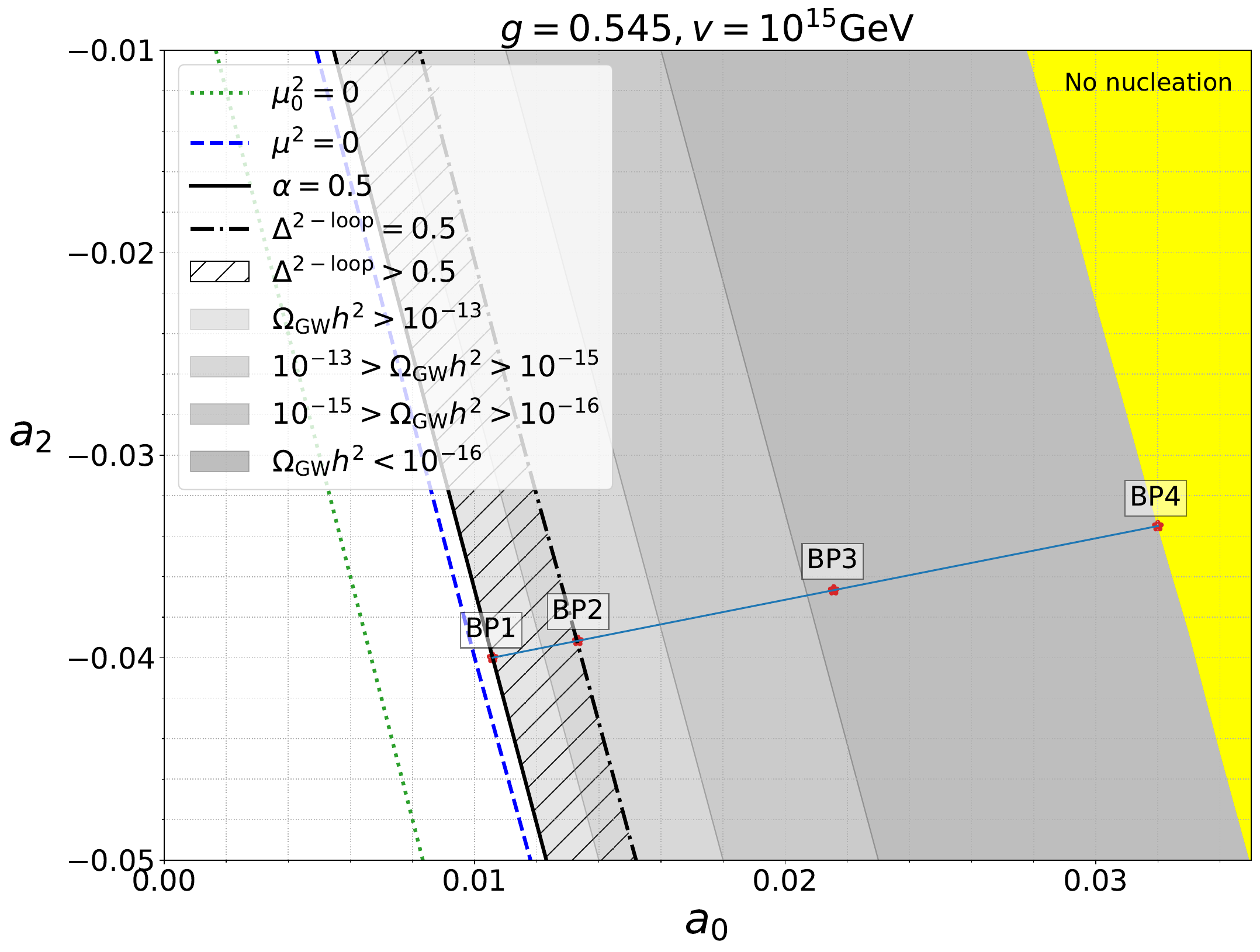}
    \caption{The same as in \fig{fig:cvgce_of_min1loop_1_1p6e16} but for
$\MG=1.0\times 10^{15}\GeV$.}
\label{fig:cvgce_of_min1loop_1p0e15}
\end{figure}

In order to see the behaviour of the observable $\Omega_\text{GW} h^2$ as a
function of both parameters $a_0$ and $a_2$, we present
\fig{fig:cvgce_of_min1loop_1_1p6e16} and
\fig{fig:cvgce_of_min1loop_1p0e15} for $\MG=1.6 \times 10^{16}\GeV$ and
$\MG=1.0\times 10^{15}\GeV$, respectively. 
We have also marked the four benchmark points (BP) as well as the lines
corresponding to the parameter values for which the curves in
\fig{fig:PTparameters} have been computed.
We show in solid gray the part of the allowed parameter space of
$a_0$ and $a_2$ that allows for a first-order phase transition.  The
shading indicates the order of magnitude of the GW signal, with lighter
shading corresponding to a stronger signal.
Inside the yellow region indicated by ``No nucleation'', there is no potential barrier and nucleation is not possible. Its exact boundary is difficult to determine because close to it the numerical algorithm eventually fails when $T_0\rightarrow T_c$.

\begin{table}
\centering
\begin{tabular}{|c|ccc|cc|cc|}
\toprule
 BP & $T_0/\MG$ & $T_c/\MG$ & $T_n/\MG$ & $\alpha$ & $\beta/H_*$ & $a_0$ & $a_2$ \\
\midrule
 BP1 & 0.0603 & 0.119 & 0.0638 & 0.50  & $7.1 \times 10^2$ & 0.0102 & $-0.04$ \\
 BP2 & 0.136  & 0.170 & 0.142  & 0.062 & $8.9 \times 10^2$ & 0.0125 & $-0.0394$ \\
 BP3 & 0.242  & 0.259 & 0.249  & 0.021 & $1.6 \times 10^3$ & 0.0202 & $-0.0377$ \\
 BP4 & 0.308  & 0.318 & 0.313  & 0.013 & $2.9 \times 10^3$ & 0.03 & $-0.0356$ \\
\bottomrule
\end{tabular}
\caption{Phase transition parameters for four benchmark points with $\MG=1.6 \times 10^{16}\,$GeV.}
\label{tbl:BP_v1016GeV}
\end{table}

We see that the borders of the shaded regions are nearly parallel.
Hence, the signal strength does not depend on $a_0$ and $a_2$ separately,
but to a good approximation only on a linear combination.  Consequently,
the two-dimensional plots in \fig{fig:PTparameters} are sufficient to
encode most of the parameter dependence, as a parallel shift of the line
connecting the benchmark points would lead to curves very similar to those 
in \fig{fig:PTparameters}.

The hatched regions mark
parameter space points where $\Delta^\text{2-loop}$ of \eq{eq:Delta2Loop}
is larger than $0.5$, indicating that the precision of our one-loop
calculation is likely to be insufficient.  As the GW signal is strongest
in these regions, extending the calculation to include higher loop
orders would be a highly desirable goal for future work.  We see that 
BP2 is at the border of the region with $\Delta^\text{2-loop}>0.5$ and
thus corresponds to the largest signal strength for which we are confident
in the accuracy of our calculation.  In contrast, BP1 has the largest
value of $\Delta^\text{2-loop}$ and should therefore be regarded with
caution.

The plots also include the lines on which the quadratic term in the
scalar potential vanishes at tree level ($\mu_0^2=0$, by which we mean
that the r.h.s.\ of \eq{eq:mu2a0a2tree} vanishes) and at the one-loop
level ($\mu^2=0$, cf.\ the discussion in Section~\ref{sec:potential}).
Evidently, a strong signal corresponds to small values of $\mu^2$.

\begin{table}
\centering
\begin{tabular}{|c|ccc|cc|cc|}
\toprule
 BP & $T_0/\MG$ & $T_c/\MG$ & $T_n/\MG$ & $\alpha$ & $\beta/H_*$ & $a_0$ & $a_2$\\
\midrule
 BP1 & 0.0603 & 0.122 & 0.0659 & 0.49  & $7.9 \times 10^2$ & 0.01058 & $-0.04$ \\
 BP2 & 0.141  & 0.176 & 0.150  & 0.058 & $1.1 \times 10^3$ & 0.0133 & $-0.0391$ \\
 BP3 & 0.244  & 0.261 & 0.252  & 0.022 & $2.5 \times 10^3$ & 0.0215 & $-0.0366$ \\
 BP4 & 0.309  & 0.319 & 0.316  & 0.013 & $5.5 \times 10^3$ & 0.032 & $-0.0335$ \\
\bottomrule
\end{tabular}
\caption{Phase transition parameters for four benchmark points with $\MG=1.0 \times 10^{15}\,$GeV.}
\label{tbl:BP_v1015GeV}
\end{table}

\section{Incomplete Phase Transitions  \label{sec:incompPT}}

In Section \ref{subsec:results}, we have seen that for both cases of $\MG=1.6 \times 10^{16}\GeV$ and $\MG=1.0 \times 10^{15}\GeV$, there is a parameter space in $a_0$ and $a_2$ (see the yellow region of \fig{fig:cvgce_of_min1loop_1_1p6e16} and \fig{fig:cvgce_of_min1loop_1p0e15}) where the phase transition does not complete at the GUT scale. For these cases, we could have the scenario ``B'', depicted in \fig{fig:possible_scale_order}, because the criterion of \eq{eq:condition_inc_FOPT} is satisfied. As pointed out in \cite{Barir:2022kzo}, the phase transition can still occur at a lower scale, leaving an imprint in the CMB\@. For this to happen the scalar field breaking the GUT theory needs to act like a spectator value during inflation, \eq{eq:GammaVLH4} has to be satisfied and $V_{\rm Inf}(\phi_{\rm Inf})\gg V(\phi)$, where we have called $\phi_{\rm Inf}$ the field triggering inflation. What is relevant for the breaking chain that we are considering is that the scalar spectrum is modified and the induced GW can be seen at lower frequencies than those of the phase transitions and the plasma. This depends on the phase transition parameters and is governed by the dimensionless parameter $\gamma_{PT}:=1/H^4 \, \Gamma/V \big(\Delta V(\phi)/\dot{\phi}_{\rm Inf} \big)^2$, defined in \cite{Barir:2022kzo}. In the yellow region of \fig{fig:cvgce_of_min1loop_1_1p6e16} and \fig{fig:cvgce_of_min1loop_1p0e15}, $1/H^4 \, \Gamma/V < 1$ and we can compute it in those regions. The other term in $\gamma_{PT}$ depends on details of inflation and since we do not have a model of inflation we can give an upper bound $\gamma_{PT}\leq 10^{-3} \big(\Delta V(\phi)/\dot{\phi}_{\rm Inf} \big)^2$, since for $V(\phi)\ll V_{\rm Inf}(\phi_{\rm Inf})$ we expect that $V(\phi)\ll V_{\rm Inf}(\phi_{\rm Inf})\ll \Mp^4$ at the end of inflation when the slow-roll parameter is close to unity. For illustration, we plot in \fig{fig:all_signals} the region in the plane of frequency vs.\ $\Omega_\text{GW} h^2$ at which these signals could potentially be observed.

\section{Gravitational Waves from the SO(10) Plasma} \label{sec:Plasma}

Any plasma in thermal equilibrium emits GW \cite{Ghiglieri:2015nfa}.
The signal is determined by the shear viscosity of the plasma.  It is guaranteed to be present in the SM, although the ultra-high frequency of
the GW makes their detection challenging.  Scenarios of physics beyond
the SM can both lower the peak frequency and enhance the signal as a result of the many additional degrees of freedom. As the shear viscosity changes during the different stages of the GUT
symmetry breaking down to the SM, we obtain a different signal during
each particular stage.  Furthermore and more importantly, as the breaking scales $M_\text{GUT}$ and $M_R$ are fixed by gauge coupling
unification, the maximum temperature of each stage is fixed, unlike the
SM, where the maximum temperature is the reheating temperature and thus
a free parameter.
In the following, we determine the GW signal for each breaking stage,
following \cite{Ghiglieri:2015nfa,Ghiglieri:2020mhm}.

In this work, we present the results for the GW signal from the plasma during the three steps of symmetry breaking shown in \eq{eq:bcsu3su2Rsu2Lu1BL} with the minimal content of matter. For each step, we take
\eq{eq:etaHTL} and use the well-known expression for the GW produced by a plasma in equilibrium \cite{Ghiglieri:2015nfa},
\begin{equation}
 \label{eq:FinalformOmegaGW}
 \Omega_\text{GW}(f,T_0) \, h^2 = \Omega_{\gamma_0} h^2
\frac{\lambda}{\Mpbar}\int^{T_{\rm Max}}_{T_{\rm EWCO}}\, dT \left(\frac{g_{*0}}{g_*(T)} \right)^{4/3} \, T^2\, \hat{k}(f,T)^3 \frac{\eta(\hat{k},T)}{\sqrt{\rho(T)}}\, \beta(T) \,,
\end{equation}
where $\Omega_{\rm {GW}}(f,T_0)$ is the fraction of energy released into GW radiation per frequency octave~\cite{Kamionkowski:1993fg}, $\lambda=30\sqrt{3}/\pi^4$ and 
\begin{equation}
    \hat{k}(f,T):=\frac{k}{T}=\left [\frac{g_{*s}(T)}{g_{*s}(T_0)} \right ]^{\frac{1}{3}}\frac{2\pi f}{T_0} \,,\quad
    f=\frac{1}{2\pi}\left [ \frac{g_{\rm *s}(T_0)}{g_{\rm *s}(T_{\rm EWCO})}\right]^{\frac{1}{3}}\left (\frac{T_0}{T_{\rm EWCO}} \right ) k_{\rm EWCO} \,.
\end{equation}
In these equations, $T_{\rm EWCO} =160\GeV$, 
\[
k_{\rm EWCO}= k(T)\left (g_{*s}(T_{\rm EWCO})/g_{*s}(T) \right )^{1/3} T_{\rm EWCO}/T
\]
is the wave number at $T$ = $T_{\rm EWCO}$,  $T_{\rm Max}$ is the maximum temperature that we identify with the reheating temperature.

The shear viscosity of the SM was computed up to the leading contribution in \cite{Ghiglieri:2015nfa}, while accounting for temperature dependence in \cite{Ghiglieri:2020mhm}. For a general theory, we have
\bea
\label{eq:eta_complete_conts}
\eta(\hat{k},T)  
&=&
 \left\{
\begin{array}{cc}
\frac{1}{8\pi}\frac{16}{g_1^4 \ln(5T/m_{D_1})} \,, & \, \hat{k} \lesssim  \alpha_1^2 \,,\\
\eta_{\rm{HTL}}(\hat{k},T)+ \eta^T(\hat{k},T) \,,
& \hat{k}  \gtrsim 3 \,,
\end{array}
\right.
\eea
where $\eta_{\rm{HTL}}$ is the Hard Thermal Logarithmic (HTL) expression~\cite{Braaten:1989mz,Ghiglieri:2020mhm}
\bea
\label{eq:etaHTL}
\eta_\text{HTL}(\hat{k},T)=\frac{1}{16 \pi} \hat{k} \, n_B(\hat{k})\, \sum^m_{n=1} d_n {\hat{m}}^2_{D_{\mathcal{G}_n}} \ln\left(4\frac{{\hat{k}}^2}{\hat{m}^2_{D_{\mathcal{G}_n}}} + 
1 \right),
\eea
where $n_B(\hat{k})=\frac{1}{e^{\hat{k}}-1}$, and  the sum is over the $m$ different groups if the model contains more than one group factor, $d_n$ is the number of generators of the given group and the normalized Debye masses are $\hat{m}^2_{D_{\mathcal{G}_n}}={m}^2_{D_{\mathcal{G}_n}}/T^2$.
The $\eta^T$ part is a function that depends on the gauge couplings at definite temperature. For the SM, this has been calculated in \cite{Ghiglieri:2020mhm}. In this work, we assume $\eta^T\rightarrow 0$ for the SO(10) model. Note that the first part in \eq{eq:eta_complete_conts} corresponds to the hydrodynamic limit (low frequency) and the second term to the leading log (high frequency) \cite{Ghiglieri:2015nfa}. 

In practice, since \eq{eq:FinalformOmegaGW} takes its dominant contribution at $T_{\rm Max}$ and since the production of the GW takes place in the radiation era, the expression for the signal reduces to 
\bea
\frac{{\Omega_{\rm {GW}}(f,T_0) h^2 }}{\Omega_{\gamma_0} h^2}&=&  \lambda\, \left[\frac{a_{\rm{Max}}T_{\rm{Max}}}{a_0 T_0}\right]^4  \frac{T_{\rm{EWCO}}-T_{\rm{Max}}}{\Mpbar}\frac{T^2_{\rm{in}}}{\sqrt{\rho}}  \, \hat{k}^3\, \eta\left(T_{\rm Max}, \hat{k}\right),
\eea
which we use to produce our plots. 

We recall that for a general gauge theory the Debye masses of a group
factor $\mathcal{G}_n$ are given by \cite{Weldon:1982bn} 
\begin{equation} \label{eq:DebyeMass}
m^2_{D_{\mathcal{G}_n}}(T) =
g^2_n(T) \, T^2 \left[
\frac{1}{3} I(\text{ad}^{(n)}) + \frac{1}{6} \sum_i I(F_i^{(n)}) + \frac{1}{6} \sum_j I(S_i^{(n)})
\right] ,
\end{equation}
where the notation for representations and Dynkin indices is explained
below \eq{eq:betagigen}.
Besides, $g_n(T)$ is the temperature-dependent gauge coupling.
From now on, we suppress the temperature dependence of the Debye masses and the gauge couplings in our formulas. 
In our model, the results for the Debye masses, calculated with the help of \texttt{GroupMath} \cite{Fonseca:2020vke},
are as follows.

\paragraph{SO(10) Plasma}
For the stage above the unification scale, we have a single group
$\mathcal{G}=SO(10)$.  There are three families of
fermions in the $\mathbf{16}$ representation, that is,
$I(F_i)=I(\rep{16})=2$ with $i=1,2,3$.
In addition, we have gauge bosons and scalars in the adjoint,
$I(\text{ad})=I(\rep{45})=8$, a real scalar multiplet with
$I(\rep{10})=1$, and complex scalars in the \rep{126}, so
$I(\rep{126}_j)=35$ for $j=1,2$.%
\footnote{Above each symmetry breaking scale, we obviously need to include all the particles that later on will be the would-be
Nambu-Goldstone bosons of the broken theory.}
Summing up these contributions, \eq{eq:DebyeMass} yields
\begin{equation}
	m^2_{D_{SO(10)}} = \frac{101}{6} \, g^2 T^2 \,,
\end{equation}
so the contribution to \eq{eq:etaHTL} is
\begin{equation}
\label{eq:SO10_Debye_Masses}
d_{SO(10)} m^2_{D_{SO(10)}} =
45 \times \frac{101}{6} g^2 T^2 \,.
\end{equation}
This value is conservative, since a complete model will include
additional contributions from multiplets needed to obtain realistic
Yukawa couplings and to accommodate Dark Matter.

\paragraph{$\mathbf{SU(3)}_C\times \mathbf{SU(2)}_L\times \mathbf{SU(2)}_R\times \mathbf{U(1)}_{B-L}$ Plasma}
Between the unification scale and the scale of $B-L$ breaking, the
plasma contains the fermions, the gauge bosons of the intermediate gauge
group $G_{3221}$, the doublet scalars originating from the \rep{10}, and
the triplet scalar breaking $SU(2)_R \times U(1)_{B-L}$ from the \rep{126}.  The representations decompose as (for the indices $a=3, 2L, 2R, B-L$)
\begin{eqnarray}
\rep{10} &=&
( \rep{1} , \rep{2} , \rep{2} , 0 ) \oplus
\text{(heavy triplets)} \,,
\nonumber\\
\rep{16} &=&
( \rep{3} , \rep{2} , \rep{1} , \tfrac{1}{3} ) \oplus
( \mathbf{\overline{3}} , \rep{1} , \rep{2} , -\tfrac{1}{3} ) \oplus
( \rep{1} , \rep{2} , \rep{1} , -1 ) \oplus
( \rep{1} , \rep{1} , \rep{2} , 1 ) \,,
\nonumber\\
\rep{126} &=&
( \rep{1} , \rep{1} , \rep{3} , 2 ) \oplus
\text{(heavy states)}
\end{eqnarray}
in terms of $G_{3221}$ multiplets, where the last number in parentheses
indicates $B-L$.  Plugging the corresponding $SU(2)$ and $SU(3)$ Dynkin
indices as well as $\frac{3}{8}(B-L)^2$ (where the factor $\frac{3}{8}$
is due to the GUT normalization of $g_{B-L}$) into \eq{eq:DebyeMass},
we find
\begin{equation}
\label{eq:G3221_Debye_Masses}
d_i m^2_{D_i}=\left\{ 
\begin{array}{c}
d_1 \frac{5}{2} g_{B-L}^2 T^2, \, d_1=1 \,, \\
\\
d_2 \, 2 g_{2L}^2 T^2, \,
d_2=3 \,, \\
\\
d_2 \frac{8}{3} g_{2R}^2 T^2, \,
d_2=3 \,, \\
\\
d_3 \, 2 g_3^2 T^2, \,
d_3=8 \,.
\end{array}
\right.
\end{equation}
In the case with $\MG=10^{15}\GeV$, the Dark Matter fermionic bi-doublet
with $B-L=0$ contributes to the $SU(2)$ group factors an additional $1/6$,
resulting in
$m^2_{D_{SU(2)_L}}=13/6 \, g^2_{2L} T^2$ and
$m^2_{D_{SU(2)_R}}=17/6 \, g^2_{2R} T^2$.

\paragraph{SM Plasma}
Below the $B-L$ breaking scale, we have the gauge group and particle
content of the SM, so we can use the known results \cite{Ghiglieri:2015nfa}%
\footnote{Taking into account the GUT normalization of $g_1$.}
\bea
\label{eq:SM_Debye_Masses}
d_i m^2_{D_i}=\left\{ 
\begin{array}{c}
d_1 \frac{11}{10} (g_1^\text{SM})^2 \, T^2, \, d_1=1 \,, \\
\\
d_2 \frac{11}{6} (g_2^\text{SM})^2 \, T^2, \, d_2=3 \,, \\
\\
d_3 2 (g_3^\text{SM})^2 \, T^2, \, d_3=8 \,.
\end{array}
\right.
\eea

\begin{figure}
    \centering
\includegraphics[width=0.7\linewidth]{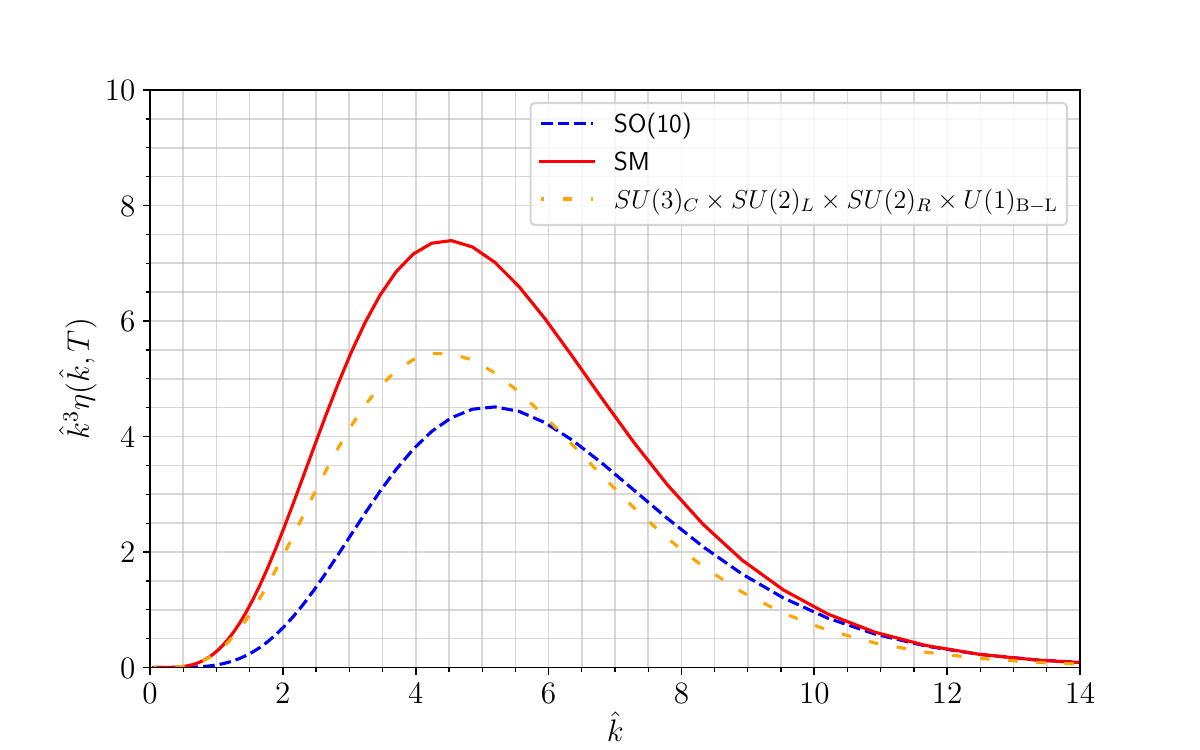}
    \caption{Comparison of the shear viscosity, $\eta$, for the case of the SM, red solid, for the $SO(10)$ case, blue long-dashed, and for the group $SU(3)_C \times SU(2)_R\times SU(2)_L\times U(1)_{B-L}$, orange short-dashed.}
    \label{fig:viscositycomp}
\end{figure}

In \fig{fig:viscositycomp} we show the comparison of the shear viscosity $\eta$ of the $SO(10)$ group, the $G_{3221}$ group and the SM group. 
We can see that the shear viscosity for $G_{3221}$ and $SO(10)$ is suppressed with respect to the SM\@.

\begin{figure}[!h]
    \centering    
    \includegraphics[width=1.0\linewidth]{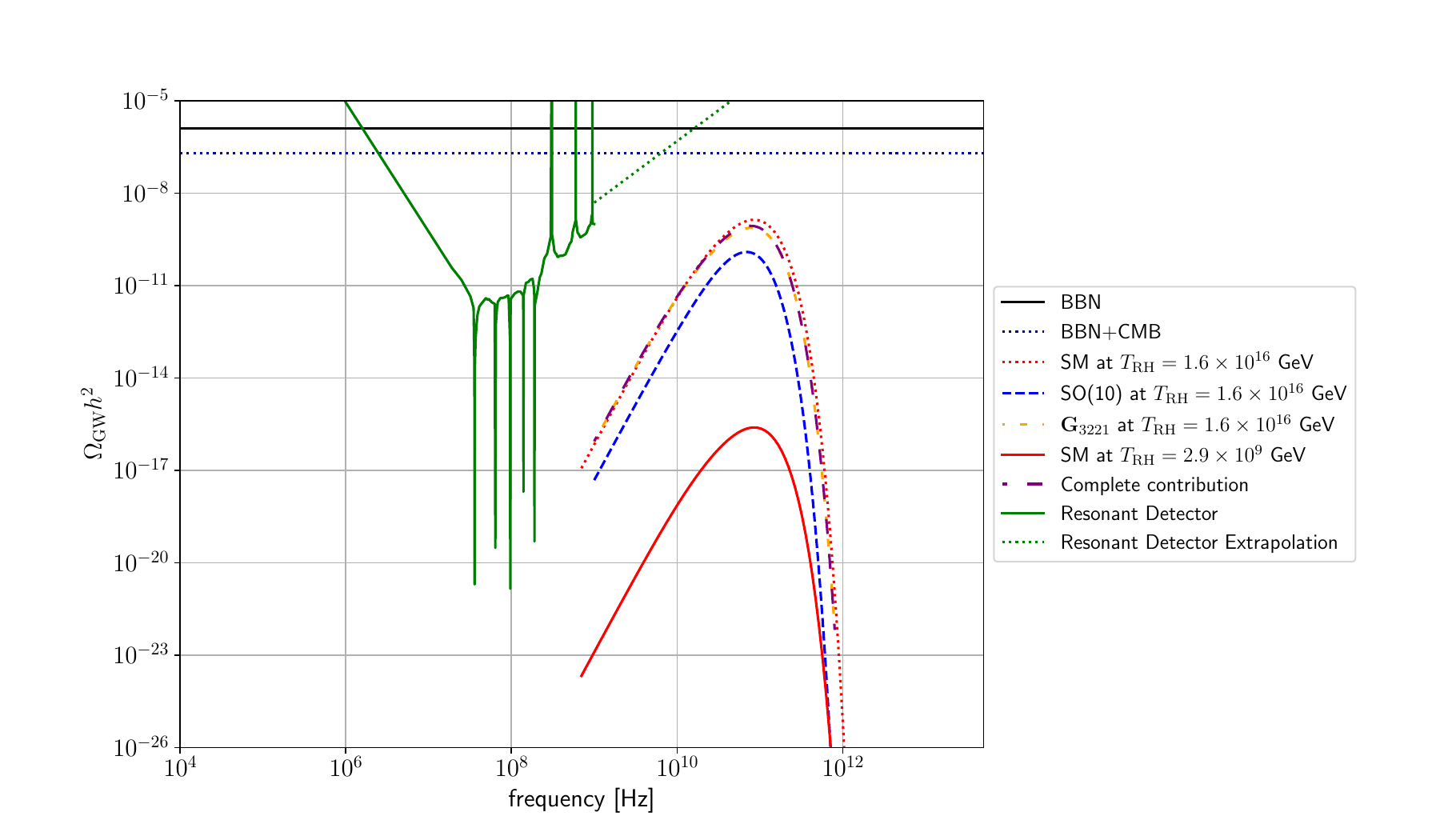}
    \includegraphics[width=1.0\linewidth]{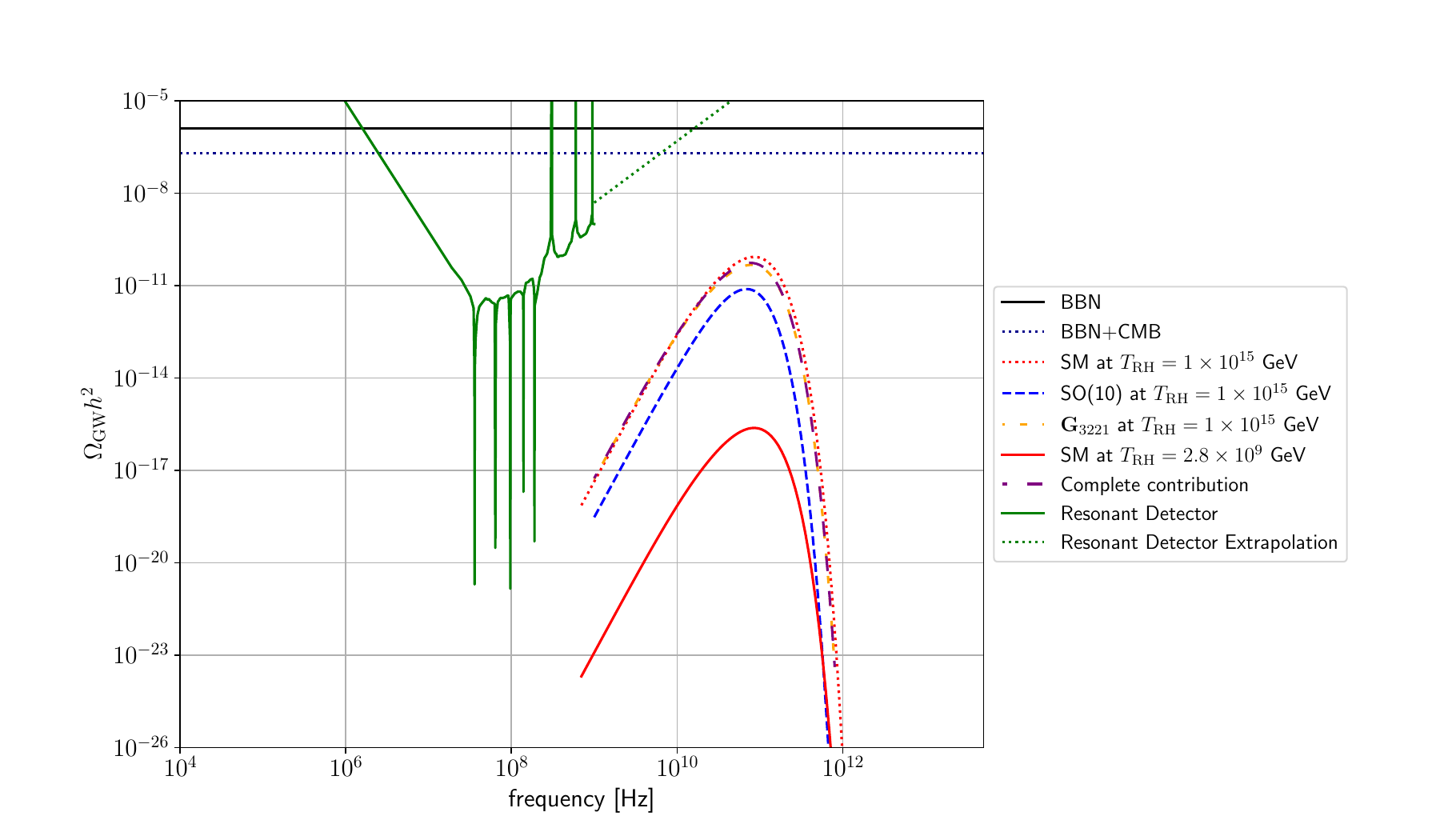}
     \caption{Contributions to GW from plasma fluctuations, shown separately for each step of the breaking chain in \eq{eq:bcsu3su2Rsu2Lu1BL}
for $\MG=1.6\times10^{16}\GeV$ (top) and $\MG=10^{15}\GeV$ (bottom). 
For comparison, we have also plotted the SM expectation for the same
reheating temperature.
``Resonant Detector'' refers to the proposal in \cite{Herman:2022fau}. In this plot we show only the leading-log part in $\eta$, \eq{eq:eta_complete_conts}, since it is the one that determines the location of the peak. In \fig{fig:all_signals} we show also the hydrodynamic limit, relevant for disentangling the signal in the low-frequency band from other sources. }
    \label{fig:GW_Plasma}
\end{figure}
Note, from \eq{eq:FinalformOmegaGW}, that the location of the
peak of the GW density, as a function of the frequency, scales as
$(g_{*0}/g_*(T))^{4/3}$ and hence the more degrees of freedom the lower
the frequency of the peak of the GW density. This is illustrated in
\fig{fig:GW_Plasma} where we have plotted each of the contributions to
the GW plasma of the breaking chain of \eq{eq:bcsu3su2Rsu2Lu1BL}. The
height of the peak is determined by the maximum attainable temperature
at each stage. Note that the $SO(10)$ group would be unbroken from $\Mp$ down
to $\MG$ but due to constraints on inflation and thermal equilibrium of
the plasma, as mentioned in Section~\ref{sec:Model}, the plasma would
enter into thermal equilibrium just close to the GUT scale and therefore
we consider the temperature of the $SO(10)$ (dashed blue line) plasma at
this temperature.  The short-dashed (orange) line represents the contribution from $G_{3221}$ and the solid red line, which is suppressed by more
than four orders of magnitude with respect to the $SO(10)$ prediction, the SM contribution. Note that the contribution from the $G_{3221}$ stage
leads the overall signal, although technically the maximum attainable
temperature is lower than for the complete $SO(10)$ group. 
For comparison, we have also plotted the GW density of the plasma of the SM at the maximum temperature of $SO(10)$ so that we can see how we could distinguish the signals for each case. An advantage of the signal predicted in $SO(10)$ compared to the SM is that, even if the height of the peak is lower, its location has a lower frequency. Furthermore, in the $SO(10)$ case, the reheating temperature is fixed by the hierarchy of the scales of the breaking chain \eqref{eq:bcsu3su2Rsu2Lu1BL}, while for the case of the SM, there is not a way to fix the reheating temperature.

\section{Conclusions}
We have investigated expected GW signals within one of the most promising breaking chains of a non-supersymmetric $SO(10)$ model \cite{Mambrini:2015vna,Graf:2016znk,Jarkovska:2021jvw,Jarkovska:2023zwv}, \eq{eq:bcsu3su2Rsu2Lu1BL}. We focused on two types of signal: the GWs produced by the high-temperature first-order phase transitions (FOPT) induced by the symmetry breaking expected in GUTs, and the stochastic background produced by the shear viscosity of the relativistic plasma in thermal equilibrium throughout the expansion history of the Universe.

As far as the FOPT is concerned, we considered a minimal particle content, including the gauge bosons, three families of fermions transforming under the representation {\bf 16} and three scalars multiplets transforming under {\bf 10}, {\bf 45} and {\bf 126}. Specifically, we analyzed the parameter space where the symmetry breaking is triggered by the {\bf 45}, whose tree-level potential is given in \eq{eq:V0tree45} and governed by two free parameters $a_0$ and $a_2$.
We studied the corresponding effective potential including the one-loop and thermal corrections, providing for the first time a compact analytic expression. We explored gauge coupling unification in this scenario, finding that it takes place between $10^{15}$ and $10^{16}\GeV$, and adopted an iterative procedure to fix the quadratic term in the potential ensuring a vev of the {\bf 45} in the correct range for each value of $a_0$ and $a_2$. To calculate the ensuing expected signals we used the parameters of the phase transition obtained from the effective potential in the well-known expressions for the sound wave \cite{Hindmarsh:2013xza,Hindmarsh:2015qta} and turbulence \cite{Caprini:2009yp,Binetruy:2012ze} contributions to the density of GW when the barrier is mainly of thermal origin. 

\begin{figure}
    \centering
    \includegraphics[width=0.98\linewidth]{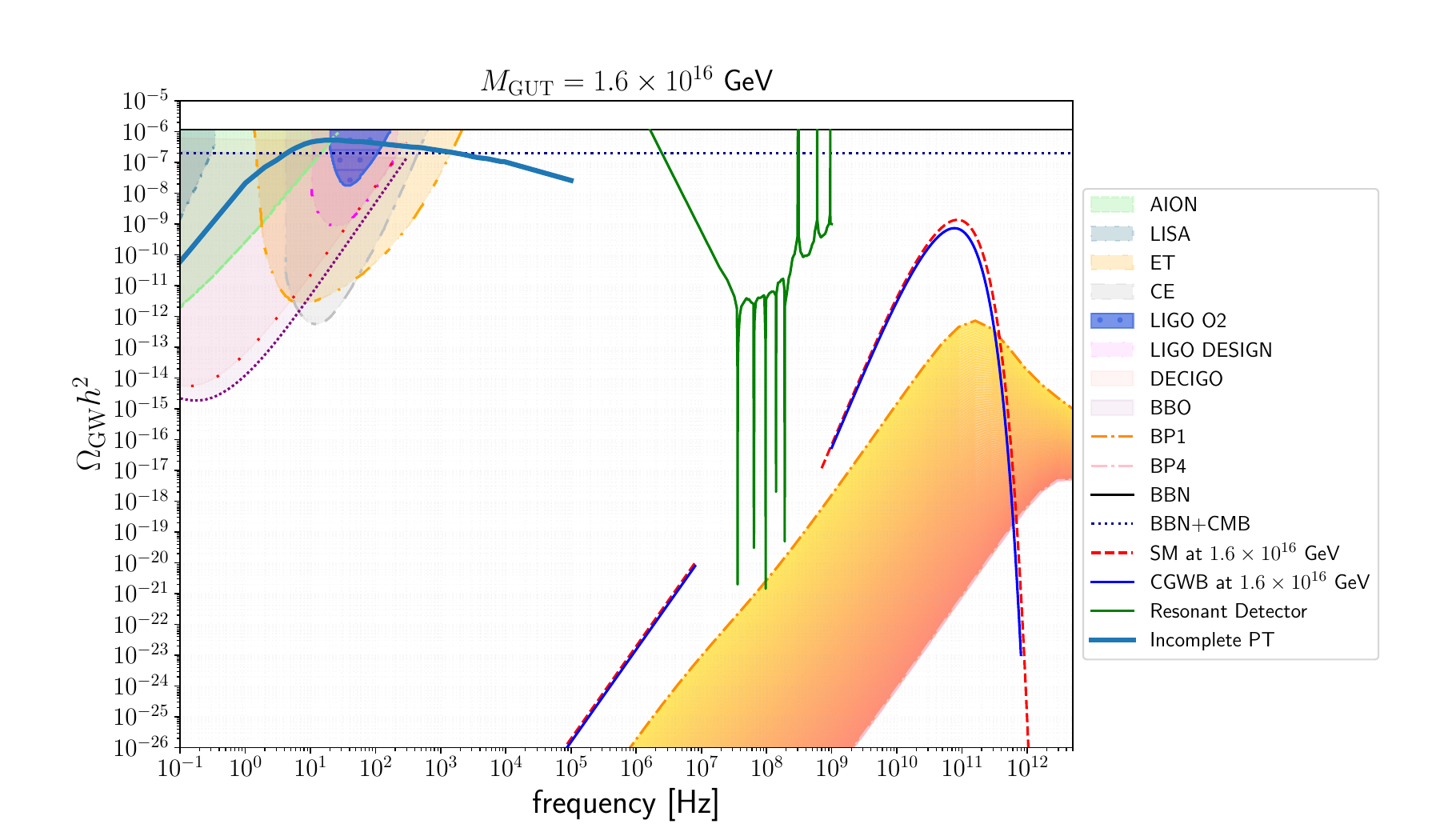}
    \includegraphics[width=0.98\linewidth]{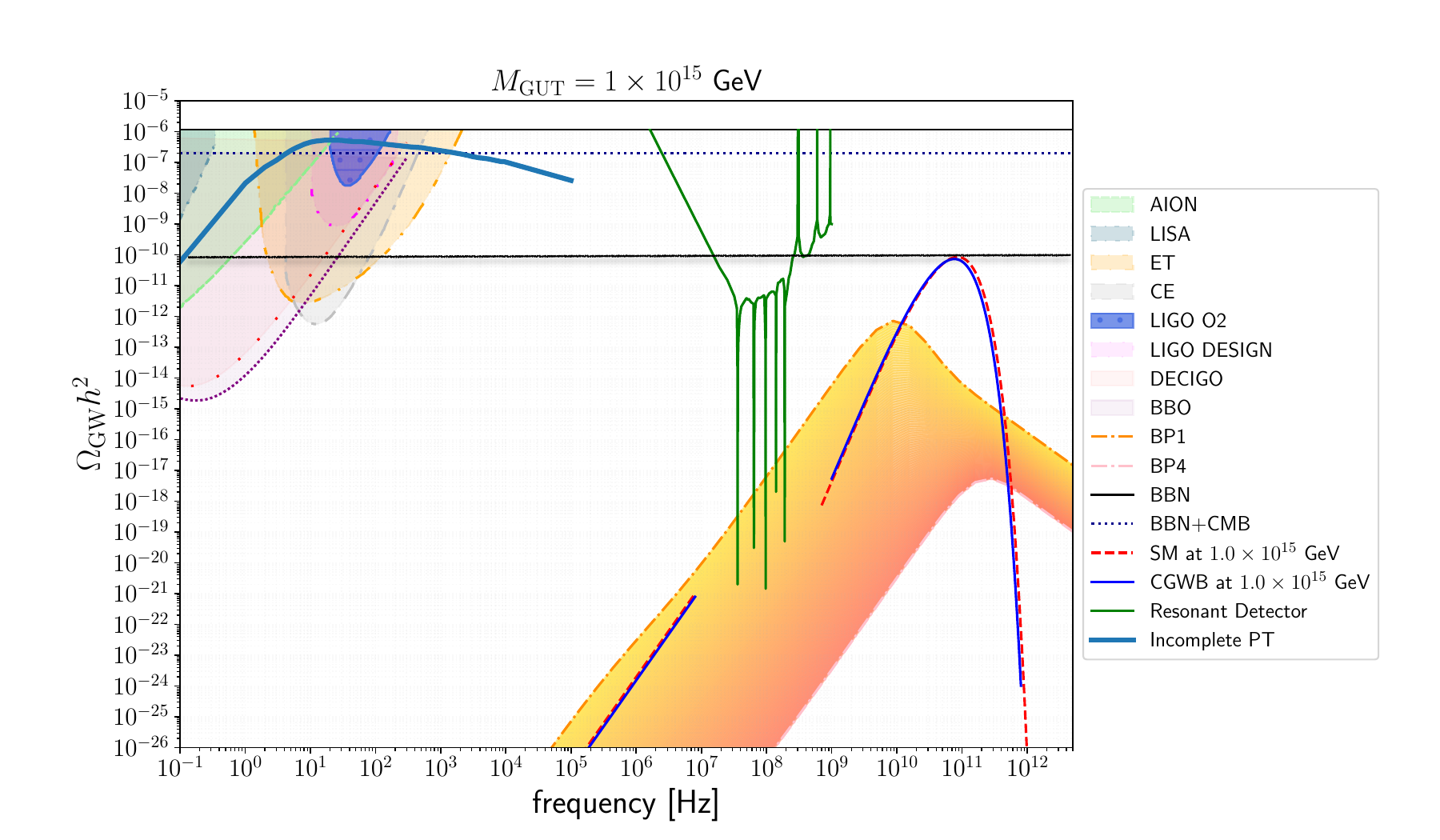}
    \caption{GW signals from the considered $SO(10)$ model for two different values of the unification scale, $\MG=1.6\times10^{16}\GeV$ (top) and $\MG=10^{15}\GeV$ (bottom). The coloured region shows the GW spectrum that can be obtained from the FOPT at the GUT scale by varying the scalar quartic couplings within the allowed region of the parameter space; the upper boundary corresponds to BP~1, while the lower boundary is obtained for BP 4.
    Solid and dashed lines show the GW produced by fluctuations in the plasma, with the $SO(10)$ prediction plotted in blue (denoted by CGWB: Cosmic GW Background), while the signal expected in the SM with the same reheating temperature is plotted in red for comparison.
    We also show the sensitivity of current and planned experiments, along with the BBN and the CMB constraints.}
    \label{fig:all_signals}
\end{figure}

We found that the FOPT takes place in a significant part of the parameter space allowed by the absence of tachyons and by proton decay. 
The GW signal peaks at frequencies of order $10^{10}$ to $10^{11}\,$Hz, depending on the unification scale, as expected for FOPT occurring at very high energy scales.
This puts it beyond the sensitivity range of pulsar timing arrays and most GW detectors but possibly within the range of proposed resonant detectors \cite{Herman:2022fau}.\footnote{
 It has been pointed out that \cite{Gehrman:2023esa,Gatti:2024mde,TitoDAgnolo:2024res}, once realistic coherence, duty-cycle and experimental losses are included, the sensitivities of these resonant detectors can be significantly decreased compared to the proposal of \cite{Herman:2022fau}, and that we include in the plots of Figs.~(\ref{fig:GW_Plasma}-\ref{fig:all_signals}).}.
We found the strongest signal in the region of the parameter space where the accuracy of the computation is limited by large contributions of higher loop orders.  Consequently, an important direction for future work is to improve the calculation of the effective potential beyond one-loop precision. We remark, that this is often also needed for models that seem to have a promising observable signal \cite{Espinosa:1996qw,Curtin:2016urg,Niemi:2024vzw}. 

Fluctuations in the primordial plasma can yield a stronger GW signal than the FOPT, at frequencies of the same order, as shown by the solid and dashed lines in \fig{fig:all_signals}. However, the $SO(10)$ signal is somewhat suppressed compared to the SM with the same reheating temperature. Consequently, observing a signal at the SM strength would in principle disfavor an interpretation in terms of the $SO(10)$ model. 
The lower-frequency part of the plotted signal corresponds to the hydrodynamic limit, i.e., the first expression in \eq{eq:eta_complete_conts}, and the high-frequency part to the leading-log result (second expression).

In \fig{fig:all_signals} we plot all GW signals that we have considered in this work, along with some current and planned experiments, in particular the resonant detector proposed in~\cite{Herman:2022fau}.
As to the possibilities of observing GW produced by the scalar power spectrum when the phase transition is not completed before inflation (see Section \ref{sec:incompPT}), the details depend on the particular inflation scenario and are therefore more model-dependent. Nevertheless, in our scenario we can put a bound on the parameter $\gamma_{PT}$ that controls the shape and strength of the signal, $\gamma_{PT}\leq 10^{-3} \big(\Delta V(\phi)/\dot{\phi}_{\rm Inf} \big)^2$. We include this limit in \fig{fig:all_signals}. All possible signals that can come from this scenario will be lower than the solid petrol blue line shown from $10^1$ to $10^4$\,Hz. 
We have not discussed GW produced by topological defects,
which have been extensively studied recently albeit in different models (see, e.g.,
\cite{Dunsky:2021tih,Chun:2021brv,Antusch:2023zjk,Buchmuller:2023aus,Lazarides:2023rqf,Fu:2023mdu,Ahmed:2023rky,King:2023wkm,Antusch:2024nqg,Maji:2024tzg,Maji:2025yms,Maji:2025thf})
and could occur at much lower frequencies. 

We have studied a minimal GUT setup that realizes a realistic symmetry
breaking chain. Although a direct solution to the Dark Matter problem is not given in this context, we have mentioned in Section \ref{sec:GCR} how the addition of a doublet of both $SU(2)_L$ and $SU(2)_R$ can provide a Dark Matter candidate, as some of us did in \cite{Biswas:2022cyh}. 
The fermion mass spectrum and perturbativity in the Higgs sector were not addressed but interestingly, both issues can be alleviated by introducing additional scalar fields at low energy. 
We have also made a number of simplifying assumptions about the mass spectrum and the contributions to the effective potential.
Finally, the potential of a FOPT during the second step of symmetry
breaking, which would produce GW with lower frequencies, warrants
exploration.
This study should therefore be considered a first step in the investigation of phase transitions in more complicated and more realistic unified models, paving the way for a promising new direction of research at the intersection of GUT model building and cosmology.

\section*{Acknowledgements}
We would like to thank Vincenzo Branchina, Michal Malinský, and Seong Chan Park for useful discussions; Rinku Maji and Qaisar Shafi for pointing out very important details of the breaking chains and useful references.
J.K.\ is supported by the \emph{Brain Pool} program of the National Research Foundation of Korea (NRF) under grant no.\ RS-2023-00283129.
The work of S.S.\ and I.J.\ is supported by the Basic Science Research Program through the NRF funded by the Ministry of Education through the Center for Quantum Spacetime (CQUeST) of Sogang University (RS-2020-NR049598).
 J.K.\ also acknowledges partial funding and hospitality from CQUeST (RS-2020-NR049598) during the final stage of this work.
The work of L.V.S.\ is supported by the NRF grant no.\ RS-2023-00273508.
I.J.\ is also partially supported by the NRF grant RS-2023-00273508.

\appendix
\section{SO(10) Group Theory \label{sec:so10}}

The generators of $SO(10)$ are conveniently labeled by antisymmetric
indices, so we denote them by $T^{\alpha\beta}=-T^{\beta\alpha}$ with
$\alpha,\beta = 1,\dots,10$.%
\footnote{We closely follow the notation of \cite{Graf:2016znk}.}
In the fundamental representation, they
are antisymmetric $10\times 10$ matrices,
\begin{equation}
	(T^{\alpha\beta})_{ij} =
	-\frac{i}{\sqrt{2}} \left(
	 \delta_{\alpha i} \delta_{\beta j} - \delta_{\alpha j} \delta_{\beta i}
	\right) ,
\end{equation}
where $i,j = 1,\dots,10$.  We do not distinguish between upper and lower
indices and imply summation over repeated indices.
The generators satisfy
\begin{equation}
	\Tr(T^{\alpha\beta} T^{\gamma\delta}) =
	\delta_{\alpha\gamma} \delta_{\beta\delta} -
	\delta_{\alpha\delta} \delta_{\beta\gamma} \,.
\end{equation}

The scalar and gauge boson fields relevant to our discussion%
\footnote{Here we do not consider the SM fermions as well as the
additional scalars that are needed for symmetry breaking but do not play
a significant role in the FOPT.}
transform under the adjoint representation \rep{45} and
are written as antisymmetric $10\times 10$ matrices as well,%
\footnote{Our expression for $A^\mu$ differs from that in
\cite{Graf:2016znk} by a factor of $\sqrt{2} i$, which ensures real and
canonically normalized components $A^\mu_{ij}$.}
\bea
	\phi &=& \frac{i}{\sqrt{2}} \phi_{\alpha\beta} T^{\alpha\beta} \,,
\\
	A^\mu &=& \frac{i}{\sqrt{2}} A^\mu_{\alpha\beta} T^{\alpha\beta} \,.
\eea
Their Lagrangian is
\begin{equation}
	\mathcal{L} =
	-\frac{1}{4} \Tr(F_{\mu\nu} F^{\mu\nu}) +
	\frac{1}{4} \Tr\bigl( (D_\mu\phi)^\dagger (D^\mu\phi) \bigr)
- V(\phi) \,,
\end{equation}
where the scalar potential is given in \eq{eq:V0tree45}.
The field-strength tensor and the covariant derivative are given by
\begin{eqnarray}
\label{eq:fieldstrength}
	F^{\mu\nu} &=&
	\partial^\nu A^\mu - \partial^\mu A^\nu - ig \left[A^\mu,A^\nu\right] ,
\\
	D^\mu\phi &=& \partial^\mu \phi - ig \left[A^\mu,\phi\right] .
\end{eqnarray}

In order to obtain the breaking pattern
$SO(10) \to SU(3)_c \times SU(2)_L \times SU(2)_R \times U(1)_{B-L}$,
we note that $SO(10)$ contains the subgroups $SO(6)$ and $SO(4)$, which
are isomorphic to $SU(4)$ and $SU(2)_L \times SU(2)_R$, respectively.
Defining the generators of $SU(4)$ as \cite{Ozer:2005dwq}
\begin{align*}
	U_1 &= -\frac{1}{\sqrt{2}} \left( T^{36} + T^{45} \right) , &
	U_8 &= \frac{1}{\sqrt{6}} \left( 2\,T^{12} - T^{34} + T^{56} \right) ,
\nonumber\\
	U_2 &= -\frac{1}{\sqrt{2}} \left( T^{46} - T^{35} \right) , &
	U_9 &= -\frac{1}{\sqrt{2}} \left( T^{14} + T^{23} \right) ,
\nonumber\\
	U_3 &= \frac{1}{\sqrt{2}} \left( T^{34} + T^{56} \right) , &
	U_{10} &= \frac{1}{\sqrt{2}} \left( T^{13} - T^{24} \right) ,
\nonumber\\
	U_4 &= \frac{1}{\sqrt{2}} \left( T^{16} + T^{25} \right) , &
	U_{11} &= \frac{1}{\sqrt{2}} \left( T^{16} - T^{25} \right) ,
\nonumber\\
	U_5 &= -\frac{1}{\sqrt{2}} \left( T^{15} - T^{26} \right) , &
	U_{12} &= \frac{1}{\sqrt{2}} \left( T^{15} + T^{26} \right) ,
\nonumber\\
	U_6 &= \frac{1}{\sqrt{2}} \left( T^{14} - T^{23} \right) , &
	U_{13} &= \frac{1}{\sqrt{2}} \left( T^{36} - T^{45} \right) ,
\nonumber\\
	U_7 &= \frac{1}{\sqrt{2}} \left( T^{13} + T^{24} \right) , &
	U_{14} &= \frac{1}{\sqrt{2}} \left( T^{35} + T^{46} \right) ,
\end{align*}
\begin{equation} \label{eq:SU4Gen}
	U_{15} = \frac{1}{\sqrt{3}} \left( T^{12} + T^{34} - T^{56} \right) ,
\end{equation}
they satisfy
\begin{equation}
	\Tr(U_a U_b) = \delta_{ab} \quad,\quad a,b=1,\dots,15 \,.
\end{equation}
The generator $U_{15}$ commutes with both the $SO(4)$ generators and
$U_{1,\dots,8}$, which generate the $SU(3)_c$ subgroup of $SU(4)$. Consequently, a vev $\braket\phi \propto U_{15}$ leads to the
desired symmetry breaking with $U(1)_{B-L}$ generated by $U_{15}$.

\section{Effective Potential \label{app:Eff_Potential}}
Based on the previous discussion of symmetry breaking at the GUT scale,
we consider a classical (or background) field proportional to the
generator $U_{15}$,
\begin{equation} \label{eq:DefPhiC}
	\phi_c = \sqrt{2}i \, \varphi_c \, U_{15} \,,
\end{equation}
where the factor $\sqrt{2}i$ ensures canonical normalization of the real
scalar field $\varphi_c$.
At the potential minimum,
\begin{equation}
	\varphi_c = v \equiv \sqrt{3} \, \omega_{BL} \,.
\end{equation}
Assuming that the symmetry breaking at the GUT scale is dominated by the
single vev~$v$, the effective potential can be approximated as a
function of the single classical field $\varphi_c$.

The gauge boson masses after SSB are obtained by evaluating the commutators
of the $SO(10)$ generators with $\braket\phi$ \cite{Graf:2016znk}.
Likewise, we find the \emph{field-dependent} mass matrix $M_g^2(\phi_c)$,
a $45\times45$ matrix with elements
\begin{equation}
	M_g^2(\phi_c)_{(\alpha\beta)(\gamma\delta)} =
	\frac{g^2}{2} \Tr\left(
	\big[ T^{(\alpha\beta)},\phi_c \big]
	\big[ T^{(\gamma\delta)},\phi_c \big] \right) ,
\end{equation}
where $(\alpha\beta)$ is an ordered index pair ($\alpha<\beta$)
determining the row and analogously $(\gamma\delta)$ is an ordered pair
determining the column.
The block of $M_g^2$ containing the masses of the $24$ gauge bosons that do
not belong to any of the subgroups in our breaking chain is diagonal%
\footnote{This changes when we add a non-zero vev breaking $SU(2)_R$.  Then, off-diagonal entries appear, which can be removed by combining pairs of gauge bosons into mass eigenstates.}
with the non-zero entries $m_{g1}^2(\phi_c)=\frac{1}{6} g^2 \varphi_c^2$,
see \tab{tab:FieldDepMasses} where we collect the field-dependent masses
of all particles relevant for our discussion.
For the masses of the $SU(4)$ gauge bosons, the corresponding block of
$M_g^2$ is not diagonal.  Hence, it is more convenient to determine
their masses in the basis of the $SU(4)$ generators, i.e., to calculate
\begin{equation}
	M_{SU(4)}^2(\phi_c)_{ab} =
	\frac{g^2}{2} \Tr\left(
	\big[ U_a,\phi_c \big]
	\big[ U_b,\phi_c \big] \right)
\end{equation}
using the $SU(4)$ generators given in Eqs.~\eqref{eq:SU4Gen}.
This yields $6$ massive gauge bosons with field-dependent masses squared
$m_{g2}^2(\phi_c) = \frac{2}{3} g^2 \varphi_c^2$.

For the scalars, the field-dependent mass-squared matrix is given by
\begin{equation}
	M^2_s(\phi_c)_{(\alpha\beta)(\gamma\delta)} =
	\left.\frac{\partial^2\,V_0(\phi)}{\partial\phi_{(\alpha\beta)}\partial\phi_{(\gamma\delta)}}
	\right|_{\phi=\phi_c} ,
\end{equation}
where $V_0(\phi)$ is the tree-level potential \eqref{eq:V0tree45}.
Evaluating the eigenvalues of this matrix, we obtain the field-dependent
scalar and NGB masses squared given in \tab{tab:FieldDepMasses}.

The thermal corrections are given by the well-known formulas \cite{PhysRevD.9.3320,Linde:1981zj}
\bea
	V_{\text{th}}(\phi_c,T) &=&
	\sum_{i=g,s,\chi} \frac{n_i}{2\pi^2}T^4 \, J_b(x)
	\quad,\quad x \equiv \frac{m_i(\phi_c)}{T} \,,
\nonumber\\
 J_b(x) &=&
 \Re \int_0^\infty dy\,y^2\ln\left[1-e^{-\sqrt{y^2+x^2}}\right], 
\label{eq:Vthermal}
\eea
where $n_i$ represents the degrees of freedom of particles.
Also, note that
we have to sum over the individual
field-dependent gauge and scalar boson masses $m_i(\phi_c)$
given in \tab{tab:FieldDepMasses}.
As we do not consider Yukawa couplings between the fermions and
the scalar $\mathbf{45}$, there is no fermionic contribution to the
thermal corrections.

\begin{figure}
    \centering
    \includegraphics[width=0.8\linewidth]{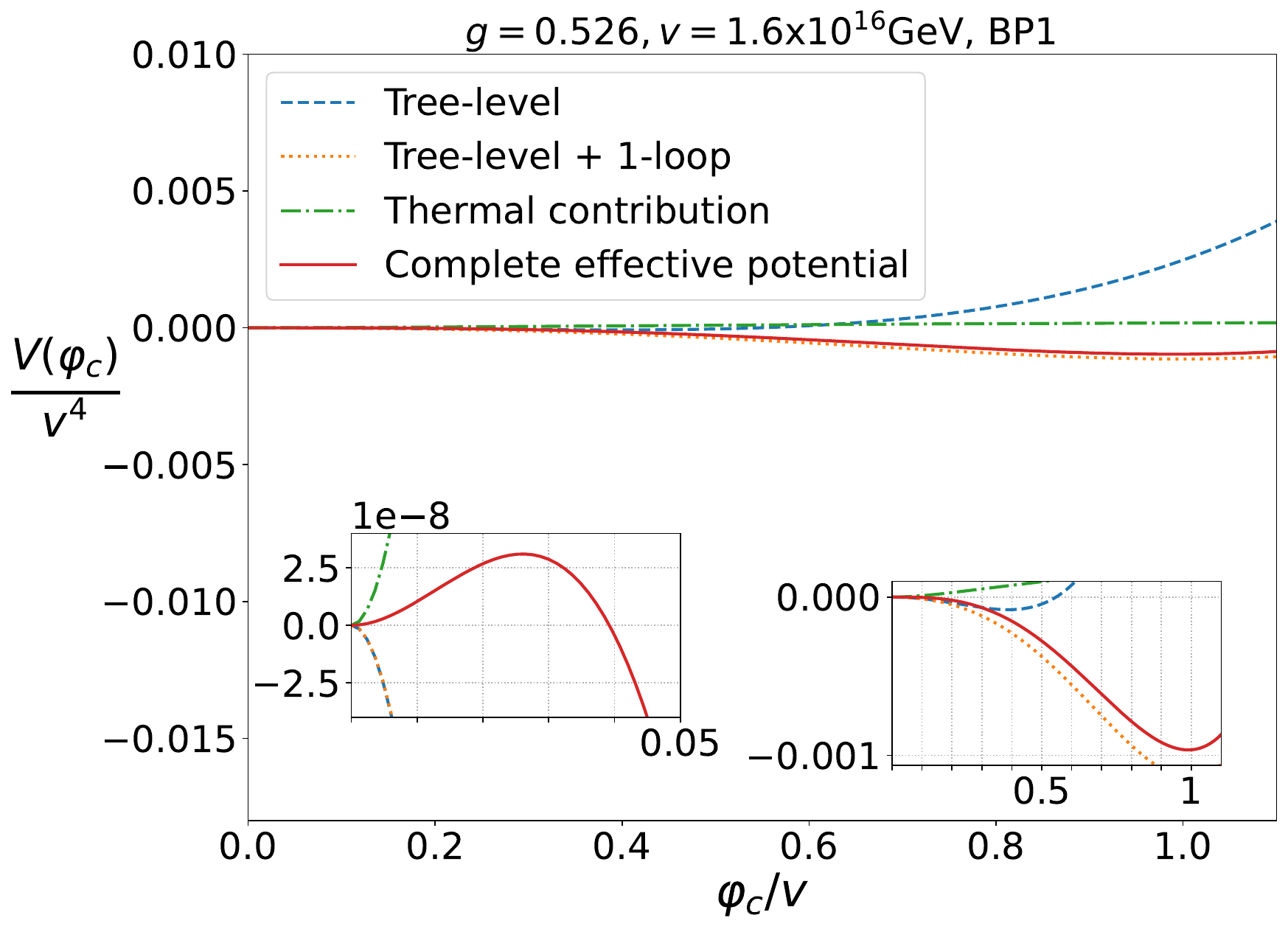}
\\ \medskip
    \includegraphics[width=0.8\linewidth]{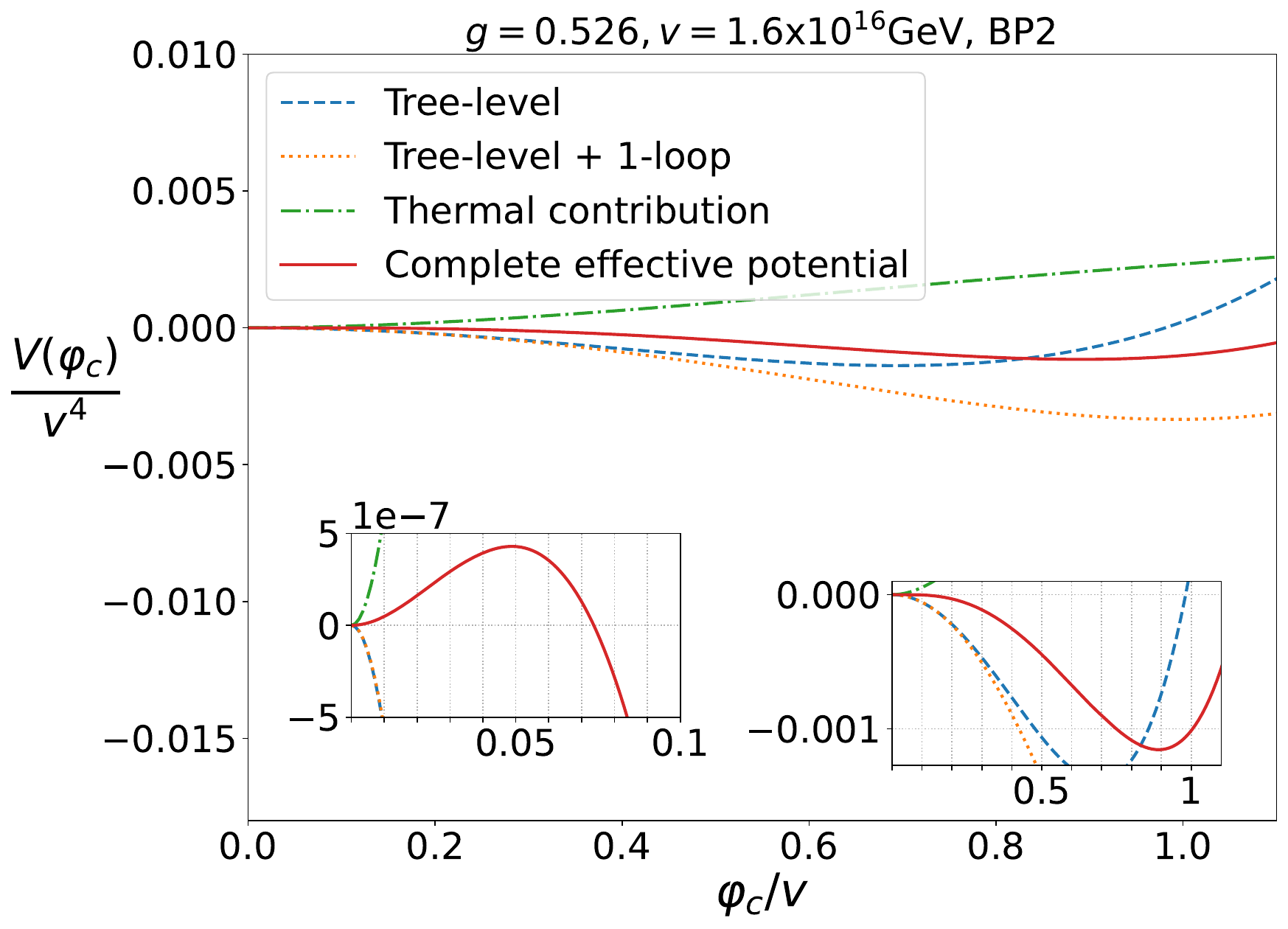}
    \caption{Effective potential for BP1 and BP2 for the case $\MG=1.6\times 10^{16}\GeV$.
	The continuous red curve represents the complete $1$-loop effective potential $V(\phi_c,T_n)$ at the nucleation temperature.
    }
    \label{fig:potential_BP2}
\end{figure}

\begin{figure}
    \centering
    \includegraphics[width=0.8\linewidth]{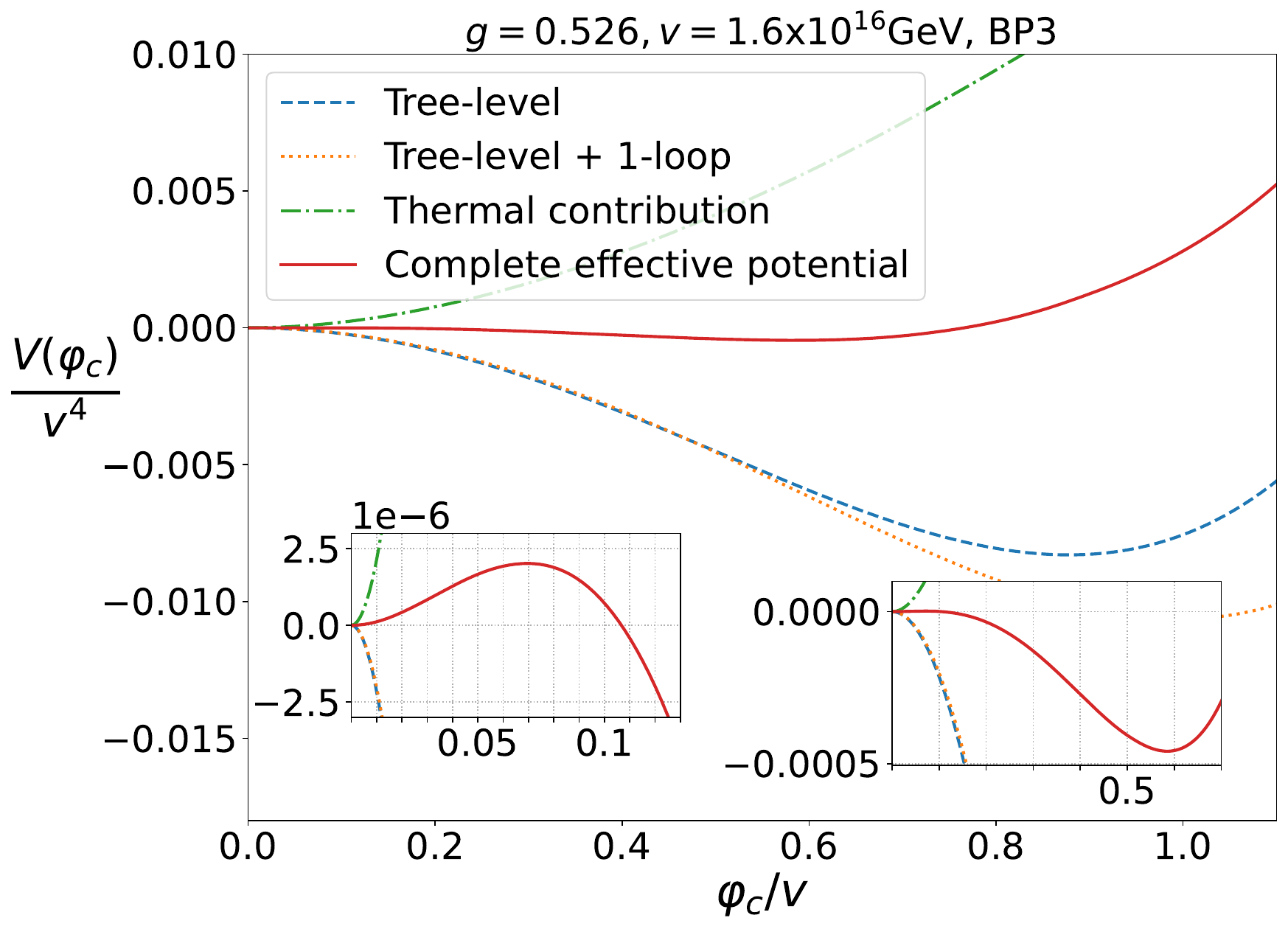}
\\ \medskip
    \includegraphics[width=0.8\linewidth]{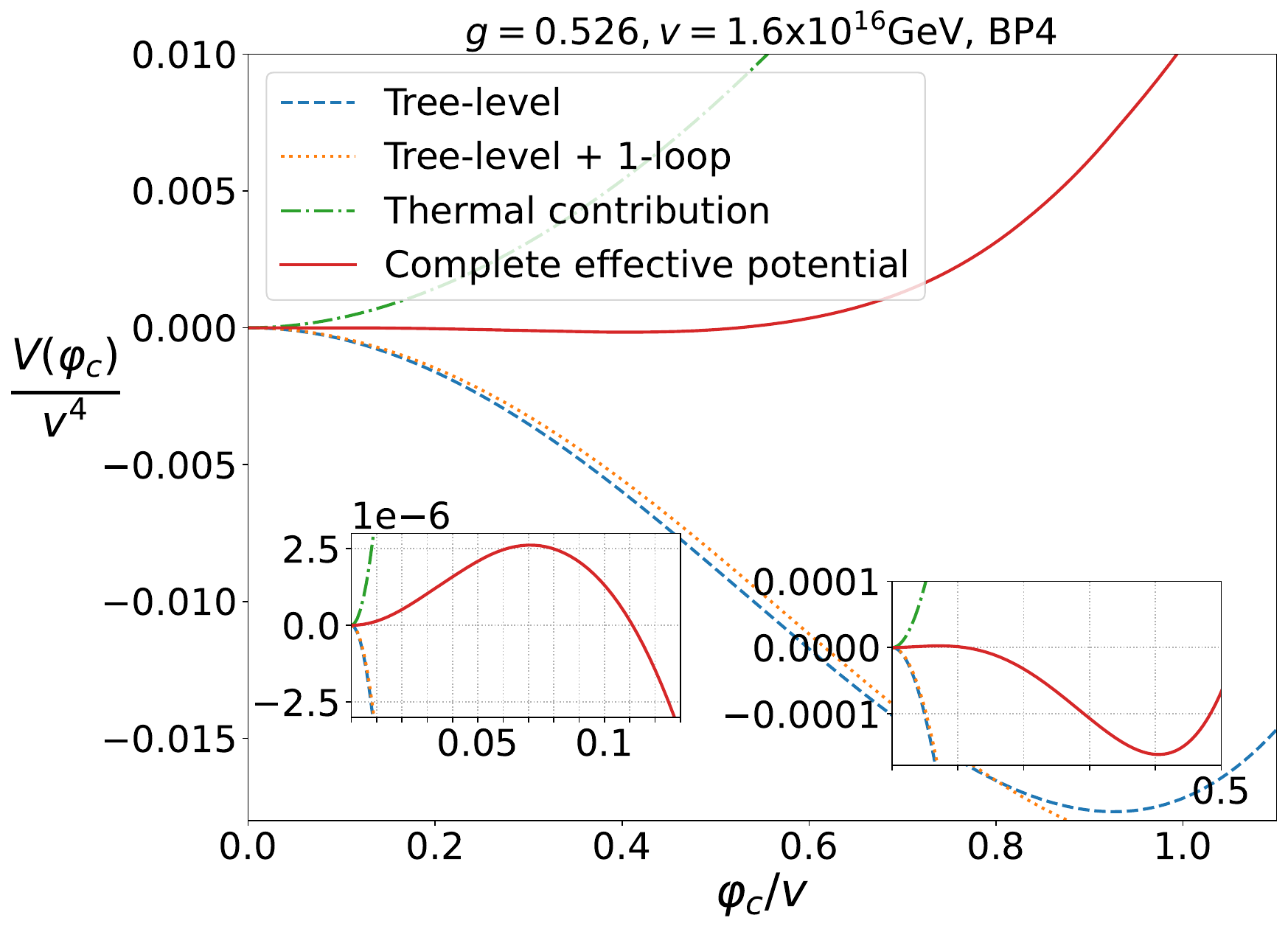}
    \caption{Same as \fig{fig:potential_BP2} for BP3 and BP4.}
    \label{fig:potential_BP4}
\end{figure}

In Figs.~\ref{fig:potential_BP2} and \ref{fig:potential_BP4}
we present plots of the effective potential for the case of
$\MG=1.6\times 10^{16}\GeV$.

\section{FOPT and GW Conventions \label{app:FOPT_conv}}
For completeness, we specify in this appendix the conventions we use for the FOPT parameters and the fits that we use for the density of GW\@. The 3D action describing the bubbles forming when the transition between the meta-stable and the true vacuum takes place is given by
\begin{equation}
\label{eq:S3T}
S_3[\phi_c(r),T] = 4 \pi \int_0^\infty r^2 dr \left[ \frac{1}{2} \left( \frac{d\phi_c(r)}{dr} \right)^2 + V(\phi_c(r),T) \right].
\end{equation}
The strength of the phase transition can be parameterized by
\begin{equation}
\label{eq:defalpha}
\alpha=\frac{1}{\rho_{\rm rad}}\left.\left[\Delta V(\phi_c,T)-\frac{T}{4}\frac{\partial\Delta V(\phi_c,T)}{\partial T}\right]\right|_{T=T_n} ,
\end{equation}
where $\Delta V(\phi_c,T)$ is the difference of the potential between the two minima (meta-stable and stable).
The $\beta$ parameter quantifies the rate at which the FOPT occurs,
\bea
\label{eq:defbeta}
\beta=H_*\left.T\frac{d}{dT}\left(\frac{S_3}{T} \right)\right|_{T=T_n\approx T_*}
.
\eea

The GW density is governed by the parameters $T_n$, $\alpha$, $\beta$, the sound speed, $c_s$, the bubble wall speed $v_w$, as well as the efficiency factors $\kappa_\nu$, and $\epsilon$. 
We employ these parameters in the well-known formulas
\cite{Hindmarsh:2013xza,Hindmarsh:2015qta} for sound wave and turbulence contributions \cite{Caprini:2009yp,Binetruy:2012ze} in the case where the barrier is mainly of thermal origin (i.e., it vanishes in the limit $T\rightarrow 0$) and the 1-loop gauge boson contributions are non-negligible.
This is known as the case of \emph{non-runaway bubbles}. 
In this case, the dominant contributions are sound waves and the magnetohydrodynamic (MHD) turbulence of the plasma.

The red-shifted sound wave contribution to the GW density observed today is
\begin{multline} \label{eq:Omegasw}
\Omega_{\rsw} h^2 (f) =
\\
2.65 \times 10^{-6} \, H_*\tau_{\rsw}
\left(\frac{\beta}{H_*}\right)^{-1} v_w
\left(\frac{\kappa_\nu \alpha }{1+\alpha }\right)^2
\left(\frac{g_*}{100}\right)^{-\frac{1}{3}}
\left(\frac{f}{f_{\rsw}}\right)^3
\left(\frac{7}{4+3 \left(f/f_{\rsw}\right)^2}\right)^{7/2} ,
\end{multline}
where
\bea
f_{\rm{sw}}=1.9\times 10^{-5} \, \frac{1}{v_w}
\left(\frac{\beta}{H_*}\right)\left(\frac{T_*  }{100 \, {\rm{GeV}}}\right)\left(\frac{g_*}{100}\right)^{1/6}{\rm{Hz}}
\eea
is the peak frequency as observed today.
The factor $\tau_{\rsw}=\min\left[\frac{1}{H_*},\frac{R_*}{\bar{U}_f}\right]$ is
the time scale of the duration of the phase transition~\cite{Ellis:2018mja,Ellis:2019oqb},
and $g_*$ and $H_*$, respectively, are the number of degrees of freedom in the thermal bath and the Hubble parameter at the time of GW production. For non-runaway bubbles, the reheating temperature and the thermal bath temperature, $T_*$, coincide with the nucleation temperature $T_n$.\footnote{Recalling that $T_{\rm rh}\approx T_n(1+\alpha)^{1/4}$ \cite{Ellis:2018mja}, this is valid only for $\alpha\ll1$.}
Thus, $\tau_{\rm sw}$ can equal either $1/H_*$ or ${R_*}/{\bar{U}_f}$, where
$H_*R_* = \max(v_w,c_s) \, (8\pi)^{1/3}(\beta/H_*)^{-1}$.
The root-mean-square (RMS) fluid velocity can be approximated as
\[
\bar{U}_f^2 \approx \frac{3}{4} \left(\frac{\kappa_\nu\alpha}{1+\alpha}\right).
\]
We emphasize that \eq{eq:Omegasw} is based on simulations that were restricted to values of $\alpha \lesssim 0.1$ and $\bar{U}_f \lesssim 0.05$ \cite{Caprini:2015zlo,Caprini:2019egz}.
The efficiency factor $\kappa_\nu$ can be approximated by \cite{Espinosa:2010hh}
\bea
\kappa_\nu \simeq \left\{ 
\begin{array}{cc}
\alpha(0.73+0.83\sqrt{\alpha} +\alpha)^{-1}, & v_w \sim 1\,,\\
v_w^{6/5} 6.9 \alpha \left( 1.36-0.037 \sqrt{\alpha} + \alpha \right)^{-1}, &  v_w \ll 1 \,.
\end{array}
\right.
\eea

The MHD turbulence provides a smaller contribution to the GW signal,
\begin{multline} \label{eq:Omegaturb}
\Omega_\text{turb} h^2 (f) =
\\
3.35 \times 10^{-4}\left(\frac{\beta}{H_*}\right)^{-1}
v_w
\left(\frac{\epsilon\, \kappa_\nu \alpha }{1+\alpha }\right)^{\frac{3}{2}}
\left(\frac{g_*}{100}\right)^{-\frac{1}{3}}
\frac{\left(f/f_{\turb}\right)^3\left(1+f/f_{\turb}\right)^{-\frac{11}{3}}}{1+8\pi\frac{f}{h_*}}
\end{multline}
with
\begin{equation}
h_*=16.5 \frac{T_*}{10^8\, \rm{GeV}}\left(\frac{g_*}{100} \right)^{1/6} \text{Hz}
\end{equation}
and the peak frequency
\bea
f_{\rm{turb}}=2.7\times 10^{-5} \, \frac{1}{v_w}\left(\frac{\beta}{H_*}\right)\left(\frac{T_*}{100\,{\rm{GeV}}}\right)\left(\frac{g_*}{100}\right)^{1/6} {\rm{Hz}} \,.
\eea
In \eq{eq:Omegaturb} we have assumed that the turbulence efficiency factor $\kappa_{\rm t}$ can be written as $\kappa_{\rm{t}}=\epsilon\kappa_\nu$,
where $\epsilon$ is another efficiency factor.
As simulations suggest that at most $5\%$ to $10\%$ of the bulk motion of the bubble wall is converted into vorticity \cite{Hindmarsh:2015qta} (which enters into the turbulence contribution), it is customary to assume a conservative value of $\epsilon=0.05$ \cite{Caprini:2015zlo}.

The bubble wall velocity $v_w$ is determined by a microphysical description of the interactions between the scalar field evolving through the bubble wall and the thermal plasma. As a precise computation is very challenging, we present the GW density profiles for the detonation velocity $v_w^d=(1/\sqrt{3}+\sqrt{\alpha^2+ 2 \alpha/3})/(1+\alpha)$ \cite{PhysRevD.25.2074}. The values $\epsilon=0.05$ and $\overline{U}_f\sim 0.05$ were obtained in the regime of detonations and small $\alpha$-deflagrations \cite{Hindmarsh:2015qta,Hindmarsh:2017gnf,Caprini:2015zlo,Caprini:2019egz}.

\section{Magnetic Flux Confinement \label{app:flux}}

As pointed out in Section~\ref{sec:Assumptions}, the symmetry breaking
at the GUT scale that we consider in our scenario leads to the production of monopoles that can overclose the Universe. In this Appendix, we provide an explicit example of how embedding $SO(10)$ into a larger group can address this feature through monopole confinement. As mentioned in the main text, this requires that the larger group is broken to $SO(10) \times U(1)_\psi$ at energies larger than the reheating temperature, so that the symmetry breaking does not produce observable topological defects.
Specifically, the $SO(10)$ monopoles remain confined (attached to cosmic strings from $U(1)_\psi$ breaking) if $U(1)_{\psi}$ is broken at the same scale and
if the monopoles carry non-vanishing $U(1)_\psi$ magnetic charge.

In order to obtain this, we first note that for our $SO(10)$ breaking chain \eqref{eq:bcsu3su2Rsu2Lu1BL},
the monopoles come from the $U(1)_{B-L}$ factor, since
they are classified via the homotopy groups \cite{nakahara}
\begin{equation}
\pi_2(Spin(10) /G_{3221}) \simeq \pi_1(G_{3221})=\pi_1(U(1)_{B-L})=\mathbb{Z}
\,.
\end{equation}
As the $\mathbb Z_4$ center of $Spin(10)$ is identified with a
subgroup of $U(1)_\psi$ in the breaking
\begin{equation}
E_6 \to \frac{Spin(10)\times U(1)_\psi}{\mathbb Z_4} \,,
\end{equation}
a monopole carries $U(1)_\psi$ magnetic charge if and only if it has
nontrivial $\mathbb{Z}_4$ center charge (see for example \cite{Lazarides:2019xai}). 
However, the $B-L$ generator is a generator of the $SU(4)$ subgroup of
$Spin(10)$, cf.~\eq{eq:SU4Gen}, and hence its monopole is
center-neutral.
One way around this issue is starting from the group $E_7$,
which possesses a fundamental weight $\Lambda_7$ that does
not lie in the root lattice $Q(E_7)$
(where $Q(G)$ represents the root lattice of the algebra $G$).
Hence, the breaking $E_7\to E_6$ (for instance via a scalar in the
$\mathbf{56}$) allows mixing between the additional $E_7$ Cartan
direction and the $\mathfrak{so}(10)$ algebra, denoted by $D_5$
\cite{hall,humphreys}.%
\footnote{Recall that the groups $SO(10)$ and $Spin(10)$ have the same algebra.}
The magnetic charge of a monopole then takes the form
\bea
m = t_{B-L} + \alpha\,\Lambda_7 \,.
\eea
As $\Lambda_7 \notin Q(E_7)$, it is not contained in the sublattice
$Q(D_5)$ either.  Thus, we have $m \notin Q(D_5)$ for $\alpha \neq 0$.
Such a magnetic charge lies in a nontrivial coset of the weight lattice relative to the $D_5$ root lattice.
This implies that
the monopole necessarily carries $U(1)_\psi$ magnetic charge and must
attach to $\psi$ strings when $U(1)_\psi$ is broken. This realizes
universal monopole-string pairing, that is, the magnetic flux is created between the monopoles in the presence of $U(1)_\psi$.
\\
Under these circumstances it only remains to ensure that the additional
field needed to break $U(1)_\psi$ does not change the phase transition that
we have studied.  This can easily be realized by choosing parameters of the
scalar potential such that the new field has only weak couplings to $\phi$
and thus no impact on the effective potential of
\eq{eq:completepot} (and induces spontaneous $U(1)_\psi$ breaking via a
cross-over or second-order phase transition).

\bibliographystyle{JHEP}
\bibliography{gw}

\providecommand{\href}[2]{#2}\begingroup\raggedright\begin{thebibliography}{10}

\bibitem{Kamionkowski:1993fg}
M.~Kamionkowski, A.~Kosowsky and M.S.~Turner, \emph{{Gravitational radiation from first order phase transitions}}, \href{https://doi.org/10.1103/PhysRevD.49.2837}{\emph{Phys. Rev. D} {\bfseries 49} (1994) 2837} [\href{https://arxiv.org/abs/astro-ph/9310044}{{\ttfamily astro-ph/9310044}}].

\bibitem{NANOGrav:2023gor}
{\scshape NANOGrav} collaboration, \emph{{The NANOGrav 15 yr Data Set: Evidence for a Gravitational-wave Background}}, \href{https://doi.org/10.3847/2041-8213/acdac6}{\emph{Astrophys. J. Lett.} {\bfseries 951} (2023) L8} [\href{https://arxiv.org/abs/2306.16213}{{\ttfamily 2306.16213}}].

\bibitem{EPTA:2023fyk}
{\scshape EPTA, InPTA} collaboration, \emph{{The second data release from the European Pulsar Timing Array -- III.~Search for gravitational wave signals}}, \href{https://doi.org/10.1051/0004-6361/202346844}{\emph{Astron. Astrophys.} {\bfseries 678} (2023) A50} [\href{https://arxiv.org/abs/2306.16214}{{\ttfamily 2306.16214}}].

\bibitem{Reardon:2023gzh}
D.J.~Reardon et~al., \emph{{Search for an Isotropic Gravitational-wave Background with the Parkes Pulsar Timing Array}}, \href{https://doi.org/10.3847/2041-8213/acdd02}{\emph{Astrophys. J. Lett.} {\bfseries 951} (2023) L6} [\href{https://arxiv.org/abs/2306.16215}{{\ttfamily 2306.16215}}].

\bibitem{Xu:2023wog}
H.~Xu et~al., \emph{{Searching for the Nano-Hertz Stochastic Gravitational Wave Background with the Chinese Pulsar Timing Array Data Release I}}, \href{https://doi.org/10.1088/1674-4527/acdfa5}{\emph{Res. Astron. Astrophys.} {\bfseries 23} (2023) 075024} [\href{https://arxiv.org/abs/2306.16216}{{\ttfamily 2306.16216}}].

\bibitem{Caldwell:2022qsj}
R.~Caldwell et~al., \emph{{Detection of early-universe gravitational-wave signatures and fundamental physics}}, \href{https://doi.org/10.1007/s10714-022-03027-x}{\emph{Gen. Rel. Grav.} {\bfseries 54} (2022) 156} [\href{https://arxiv.org/abs/2203.07972}{{\ttfamily 2203.07972}}].

\bibitem{Arzoumanian:2020vkk}
{\scshape NANOGrav} collaboration, \emph{{The NANOGrav 12.5 yr Data Set: Search for an Isotropic Stochastic Gravitational-wave Background}}, \href{https://doi.org/10.3847/2041-8213/abd401}{\emph{Astrophys. J. Lett.} {\bfseries 905} (2020) L34} [\href{https://arxiv.org/abs/2009.04496}{{\ttfamily 2009.04496}}].

\bibitem{Kume:2024xve}
J.~Kume and M.~Hindmarsh, \emph{{Revised bounds on local cosmic strings from NANOGrav observations}}, \href{https://doi.org/10.1088/1475-7516/2024/12/001}{\emph{JCAP} {\bfseries 12} (2024) 001} [\href{https://arxiv.org/abs/2404.02705}{{\ttfamily 2404.02705}}].

\bibitem{Dunsky:2021tih}
D.I.~Dunsky, A.~Ghoshal, H.~Murayama, Y.~Sakakihara and G.~White, \emph{{GUTs, hybrid topological defects, and gravitational waves}}, \href{https://doi.org/10.1103/PhysRevD.106.075030}{\emph{Phys. Rev. D} {\bfseries 106} (2022) 075030} [\href{https://arxiv.org/abs/2111.08750}{{\ttfamily 2111.08750}}].

\bibitem{Chun:2021brv}
E.J.~Chun and L.~Velasco-Sevilla, \emph{{Tracking down the route to the SM with inflation and gravitational waves}}, \href{https://doi.org/10.1103/PhysRevD.106.035008}{\emph{Phys. Rev. D} {\bfseries 106} (2022) 035008} [\href{https://arxiv.org/abs/2112.14483}{{\ttfamily 2112.14483}}].

\bibitem{Huang:2017laj}
F.P.~Huang and X.~Zhang, \emph{{Probing the gauge symmetry breaking of the early universe in 3-3-1 models and beyond by gravitational waves}}, \href{https://doi.org/10.1016/j.physletb.2018.11.024}{\emph{Phys. Lett. B} {\bfseries 788} (2019) 288} [\href{https://arxiv.org/abs/1701.04338}{{\ttfamily 1701.04338}}].

\bibitem{Croon:2018kqn}
D.~Croon, T.E.~Gonzalo and G.~White, \emph{{Gravitational Waves from a Pati-Salam Phase Transition}}, \href{https://doi.org/10.1007/JHEP02(2019)083}{\emph{JHEP} {\bfseries 02} (2019) 083} [\href{https://arxiv.org/abs/1812.02747}{{\ttfamily 1812.02747}}].

\bibitem{Brdar:2019fur}
V.~Brdar, L.~Graf, A.J.~Helmboldt and X.-J.~Xu, \emph{{Gravitational Waves as a Probe of Left-Right Symmetry Breaking}}, \href{https://doi.org/10.1088/1475-7516/2019/12/027}{\emph{JCAP} {\bfseries 12} (2019) 027} [\href{https://arxiv.org/abs/1909.02018}{{\ttfamily 1909.02018}}].

\bibitem{Graf:2021xku}
L.~Gr\'af, S.~Jana, A.~Kaladharan and S.~Saad, \emph{{Gravitational wave imprints of left-right symmetric model with minimal Higgs sector}}, \href{https://doi.org/10.1088/1475-7516/2022/05/003}{\emph{JCAP} {\bfseries 05} (2022) 003} [\href{https://arxiv.org/abs/2112.12041}{{\ttfamily 2112.12041}}].

\bibitem{Barir:2022kzo}
J.~Barir, M.~Geller, C.~Sun and T.~Volansky, \emph{{Gravitational waves from incomplete inflationary phase transitions}}, \href{https://doi.org/10.1103/PhysRevD.108.115016}{\emph{Phys. Rev. D} {\bfseries 108} (2023) 115016} [\href{https://arxiv.org/abs/2203.00693}{{\ttfamily 2203.00693}}].

\bibitem{Lazarides:1980cc}
G.~Lazarides, M.~Magg and Q.~Shafi, \emph{{Phase Transitions and Magnetic Monopoles in SO(10)}}, \href{https://doi.org/10.1016/0370-2693(80)90553-5}{\emph{Phys. Lett. B} {\bfseries 97} (1980) 87}.

\bibitem{Kibble:1982dd}
T.W.B.~Kibble, G.~Lazarides and Q.~Shafi, \emph{{Walls Bounded by Strings}}, \href{https://doi.org/10.1103/PhysRevD.26.435}{\emph{Phys. Rev. D} {\bfseries 26} (1982) 435}.

\bibitem{Jeannerot:2003qv}
R.~Jeannerot, J.~Rocher and M.~Sakellariadou, \emph{{How generic is cosmic string formation in SUSY GUTs}}, \href{https://doi.org/10.1103/PhysRevD.68.103514}{\emph{Phys. Rev. D} {\bfseries 68} (2003) 103514} [\href{https://arxiv.org/abs/hep-ph/0308134}{{\ttfamily hep-ph/0308134}}].

\bibitem{Maji:2025yms}
R.~Maji and Q.~Shafi, \emph{{C-parity, magnetic monopoles, and higher frequency gravitational waves}}, \href{https://doi.org/10.1103/PhysRevD.111.075027}{\emph{Phys. Rev. D} {\bfseries 111} (2025) 075027} [\href{https://arxiv.org/abs/2502.10135}{{\ttfamily 2502.10135}}].

\bibitem{Bertolini:2009es}
S.~Bertolini, L.~Di~Luzio and M.~Malinsky, \emph{{On the vacuum of the minimal nonsupersymmetric SO(10) unification}}, \href{https://doi.org/10.1103/PhysRevD.81.035015}{\emph{Phys. Rev. D} {\bfseries 81} (2010) 035015} [\href{https://arxiv.org/abs/0912.1796}{{\ttfamily 0912.1796}}].

\bibitem{Bertolini:2012im}
S.~Bertolini, L.~Di~Luzio and M.~Malinsky, \emph{{Seesaw Scale in the Minimal Renormalizable SO(10) Grand Unification}}, \href{https://doi.org/10.1103/PhysRevD.85.095014}{\emph{Phys. Rev. D} {\bfseries 85} (2012) 095014} [\href{https://arxiv.org/abs/1202.0807}{{\ttfamily 1202.0807}}].

\bibitem{Graf:2016znk}
L.~Gr\'af, M.~Malinsk\'y, T.~Mede and V.~Susi\v{c}, \emph{{One-loop pseudo-Goldstone masses in the minimal $SO(10)$ Higgs model}}, \href{https://doi.org/10.1103/PhysRevD.95.075007}{\emph{Phys. Rev. D} {\bfseries 95} (2017) 075007} [\href{https://arxiv.org/abs/1611.01021}{{\ttfamily 1611.01021}}].

\bibitem{Jarkovska:2021jvw}
K.~Jarkovsk\'a, M.~Malinsk\'y, T.~Mede and V.~Susi\v{c}, \emph{{Quantum nature of the minimal potentially realistic SO(10) Higgs model}}, \href{https://doi.org/10.1103/PhysRevD.105.095003}{\emph{Phys. Rev. D} {\bfseries 105} (2022) 095003} [\href{https://arxiv.org/abs/2109.06784}{{\ttfamily 2109.06784}}].

\bibitem{Jarkovska:2023zwv}
K.~Jarkovsk\'a, M.~Malinsk\'y and V.~Susi\v{c}, \emph{{Trouble with the minimal renormalizable SO(10) GUT}}, \href{https://doi.org/10.1103/PhysRevD.108.055003}{\emph{Phys. Rev. D} {\bfseries 108} (2023) 055003} [\href{https://arxiv.org/abs/2304.14227}{{\ttfamily 2304.14227}}].

\bibitem{Vilenkin:1982hm}
A.~Vilenkin, \emph{Cosmological evolution of monopoles connected by strings}, \href{https://doi.org/10.1016/0550-3213(82)90037-2}{\emph{Nucl. Phys. B} {\bfseries 196} (1982) 240}.

\bibitem{Hindmarsh:1985xc}
M.~Hindmarsh and T.W.B.~Kibble, \emph{{Monopoles on Strings}}, \href{https://doi.org/10.1103/PhysRevLett.55.2398}{\emph{Phys. Rev. Lett.} {\bfseries 55} (1985) 2398}.

\bibitem{Preskill:1992ck}
J.~Preskill and A.~Vilenkin, \emph{{Decay of metastable topological defects}}, \href{https://doi.org/10.1103/PhysRevD.47.2324}{\emph{Phys. Rev. D} {\bfseries 47} (1993) 2324} [\href{https://arxiv.org/abs/hep-ph/9209210}{{\ttfamily hep-ph/9209210}}].

\bibitem{Blanco-Pillado:2007kxw}
J.J.~Blanco-Pillado and K.D.~Olum, \emph{{Monopole annihilation in cosmic necklaces}}, \href{https://doi.org/10.1088/1475-7516/2010/05/014}{\emph{JCAP} {\bfseries 05} (2010) 014} [\href{https://arxiv.org/abs/0707.3460}{{\ttfamily 0707.3460}}].

\bibitem{Kibble:2015twa}
T.W.B.~Kibble and T.~Vachaspati, \emph{{Monopoles on strings}}, \href{https://doi.org/10.1088/0954-3899/42/9/094002}{\emph{J. Phys. G} {\bfseries 42} (2015) 094002} [\href{https://arxiv.org/abs/1506.02022}{{\ttfamily 1506.02022}}].

\bibitem{Lazarides:2019xai}
G.~Lazarides and Q.~Shafi, \emph{{Monopoles, Strings, and Necklaces in $SO(10)$ and $E_6$}}, \href{https://doi.org/10.1007/JHEP10(2019)193}{\emph{JHEP} {\bfseries 10} (2019) 193} [\href{https://arxiv.org/abs/1904.06880}{{\ttfamily 1904.06880}}].

\bibitem{Lyth:1995ka}
D.H.~Lyth and E.D.~Stewart, \emph{{Thermal inflation and the moduli problem}}, \href{https://doi.org/10.1103/PhysRevD.53.1784}{\emph{Phys. Rev. D} {\bfseries 53} (1996) 1784} [\href{https://arxiv.org/abs/hep-ph/9510204}{{\ttfamily hep-ph/9510204}}].

\bibitem{Lyth:1995hj}
D.H.~Lyth and E.D.~Stewart, \emph{{Cosmology with a TeV mass GUT Higgs}}, \href{https://doi.org/10.1103/PhysRevLett.75.201}{\emph{Phys. Rev. Lett.} {\bfseries 75} (1995) 201} [\href{https://arxiv.org/abs/hep-ph/9502417}{{\ttfamily hep-ph/9502417}}].

\bibitem{Hu:2025xdt}
X.-H.~Hu and Y.-L.~Zhou, \emph{{Gravitational waves of GUT phase transition during inflation}}, \href{https://doi.org/10.1103/kbzd-kgdr}{\emph{Phys. Rev. D} {\bfseries 111} (2025) 115003} [\href{https://arxiv.org/abs/2501.01491}{{\ttfamily 2501.01491}}].

\bibitem{Vachaspati:1992fi}
T.~Vachaspati and R.~Watkins, \emph{{Bound states can stabilize electroweak strings}}, \href{https://doi.org/10.1016/0550-3213(93)90207-Z}{\emph{Nucl. Phys. B} {\bfseries 397} (1993) 648} [\href{https://arxiv.org/abs/hep-ph/9211284}{{\ttfamily hep-ph/9211284}}].

\bibitem{Achucarro:1999it}
A.~Achucarro and T.~Vachaspati, \emph{{Semilocal and electroweak strings}}, \href{https://doi.org/10.1016/S0370-1573(99)00103-9}{\emph{Phys. Rept.} {\bfseries 327} (2000) 347} [\href{https://arxiv.org/abs/hep-ph/9904229}{{\ttfamily hep-ph/9904229}}].

\bibitem{Planck:2018jri}
{\scshape Planck} collaboration, \emph{{Planck 2018 results. X. Constraints on inflation}}, \href{https://doi.org/10.1051/0004-6361/201833887}{\emph{Astron. Astrophys.} {\bfseries 641} (2020) A10} [\href{https://arxiv.org/abs/1807.06211}{{\ttfamily 1807.06211}}].

\bibitem{Kolb:1990vq}
E.W.~Kolb and M.S.~Turner, \emph{{The Early Universe}}, Taylor and Francis (2019), \href{https://doi.org/10.1201/9780429492860}{10.1201/9780429492860}.

\bibitem{Mambrini:2015vna}
Y.~Mambrini, N.~Nagata, K.A.~Olive, J.~Quevillon and J.~Zheng, \emph{{Dark matter and gauge coupling unification in nonsupersymmetric SO(10) grand unified models}}, \href{https://doi.org/10.1103/PhysRevD.91.095010}{\emph{Phys. Rev. D} {\bfseries 91} (2015) 095010} [\href{https://arxiv.org/abs/1502.06929}{{\ttfamily 1502.06929}}].

\bibitem{Biswas:2022cyh}
A.~Biswas, A.~Kar, H.~Kim, S.~Scopel and L.~Velasco-Sevilla, \emph{{Improved white dwarves constraints on inelastic dark matter and left-right symmetric models}}, \href{https://doi.org/10.1103/PhysRevD.106.083012}{\emph{Phys. Rev. D} {\bfseries 106} (2022) 083012} [\href{https://arxiv.org/abs/2206.06667}{{\ttfamily 2206.06667}}].

\bibitem{Buras:1998raa}
A.J.~Buras, \emph{{Weak Hamiltonian, CP violation and rare decays}},  in \emph{{Les Houches Summer School in Theoretical Physics, Session 68: Probing the Standard Model of Particle Interactions}}, pp.~281--539, 1998 [\href{https://arxiv.org/abs/hep-ph/9806471}{{\ttfamily hep-ph/9806471}}].

\bibitem{Plehn:2009nd}
T.~Plehn, \emph{{Lectures on LHC Physics}}, \href{https://doi.org/10.1007/978-3-642-24040-9}{\emph{Lect. Notes Phys.} {\bfseries 844} (2012) 1} [\href{https://arxiv.org/abs/0910.4182}{{\ttfamily 0910.4182}}].

\bibitem{Chakrabortty:2019fov}
J.~Chakrabortty, R.~Maji and S.F.~King, \emph{{Unification, Proton Decay and Topological Defects in non-SUSY GUTs with Thresholds}}, \href{https://doi.org/10.1103/PhysRevD.99.095008}{\emph{Phys. Rev. D} {\bfseries 99} (2019) 095008} [\href{https://arxiv.org/abs/1901.05867}{{\ttfamily 1901.05867}}].

\bibitem{Machacek:1984zw}
M.E.~Machacek and M.T.~Vaughn, \emph{{Two Loop Renormalization Group Equations in a General Quantum Field Theory. 3. Scalar Quartic Couplings}}, \href{https://doi.org/10.1016/0550-3213(85)90040-9}{\emph{Nucl. Phys. B} {\bfseries 249} (1985) 70}.

\bibitem{Ellis:2019fwf}
J.~Ellis, J.L.~Evans, N.~Nagata, K.A.~Olive and L.~Velasco-Sevilla, \emph{{Supersymmetric proton decay revisited}}, \href{https://doi.org/10.1140/epjc/s10052-020-7872-3}{\emph{Eur. Phys. J. C} {\bfseries 80} (2020) 332} [\href{https://arxiv.org/abs/1912.04888}{{\ttfamily 1912.04888}}].

\bibitem{Super-Kamiokande:2020wjk}
{\scshape Super-Kamiokande} collaboration, \emph{{Search for proton decay via $p\to e^+\pi^0$ and $p\to \mu^+\pi^0$ with an enlarged fiducial volume in Super-Kamiokande I-IV}}, \href{https://doi.org/10.1103/PhysRevD.102.112011}{\emph{Phys. Rev. D} {\bfseries 102} (2020) 112011} [\href{https://arxiv.org/abs/2010.16098}{{\ttfamily 2010.16098}}].

\bibitem{Hyper-Kamiokande:2018ofw}
{\scshape Hyper-Kamiokande} collaboration, \emph{{Hyper-Kamiokande Design Report}},  \href{https://arxiv.org/abs/1805.04163}{{\ttfamily 1805.04163}}.

\bibitem{Hindmarsh:2020hop}
M.B.~Hindmarsh, M.~L\"uben, J.~Lumma and M.~Pauly, \emph{{Phase transitions in the early universe}}, \href{https://doi.org/10.21468/SciPostPhysLectNotes.24}{\emph{SciPost Phys. Lect. Notes} {\bfseries 24} (2021) 1} [\href{https://arxiv.org/abs/2008.09136}{{\ttfamily 2008.09136}}].

\bibitem{Huber:2008hg}
S.J.~Huber and T.~Konstandin, \emph{{Gravitational Wave Production by Collisions: More Bubbles}}, \href{https://doi.org/10.1088/1475-7516/2008/09/022}{\emph{JCAP} {\bfseries 09} (2008) 022} [\href{https://arxiv.org/abs/0806.1828}{{\ttfamily 0806.1828}}].

\bibitem{Hindmarsh:2013xza}
M.~Hindmarsh, S.J.~Huber, K.~Rummukainen and D.J.~Weir, \emph{{Gravitational waves from the sound of a first order phase transition}}, \href{https://doi.org/10.1103/PhysRevLett.112.041301}{\emph{Phys. Rev. Lett.} {\bfseries 112} (2014) 041301} [\href{https://arxiv.org/abs/1304.2433}{{\ttfamily 1304.2433}}].

\bibitem{Hindmarsh:2015qta}
M.~Hindmarsh, S.J.~Huber, K.~Rummukainen and D.J.~Weir, \emph{{Numerical simulations of acoustically generated gravitational waves at a first order phase transition}}, \href{https://doi.org/10.1103/PhysRevD.92.123009}{\emph{Phys. Rev. D} {\bfseries 92} (2015) 123009} [\href{https://arxiv.org/abs/1504.03291}{{\ttfamily 1504.03291}}].

\bibitem{Kosowsky:2001xp}
A.~Kosowsky, A.~Mack and T.~Kahniashvili, \emph{{Gravitational radiation from cosmological turbulence}}, \href{https://doi.org/10.1103/PhysRevD.66.024030}{\emph{Phys. Rev. D} {\bfseries 66} (2002) 024030} [\href{https://arxiv.org/abs/astro-ph/0111483}{{\ttfamily astro-ph/0111483}}].

\bibitem{Nicolis:2003tg}
A.~Nicolis, \emph{{Relic gravitational waves from colliding bubbles and cosmic turbulence}}, \href{https://doi.org/10.1088/0264-9381/21/4/L05}{\emph{Class. Quant. Grav.} {\bfseries 21} (2004) L27} [\href{https://arxiv.org/abs/gr-qc/0303084}{{\ttfamily gr-qc/0303084}}].

\bibitem{Caprini:2006jb}
C.~Caprini and R.~Durrer, \emph{{Gravitational waves from stochastic relativistic sources: Primordial turbulence and magnetic fields}}, \href{https://doi.org/10.1103/PhysRevD.74.063521}{\emph{Phys. Rev. D} {\bfseries 74} (2006) 063521} [\href{https://arxiv.org/abs/astro-ph/0603476}{{\ttfamily astro-ph/0603476}}].

\bibitem{PhysRevD.86.103005}
T.~Kahniashvili, A.~Brandenburg, L.~Campanelli, B.~Ratra and A.G.~Tevzadze, \emph{{Evolution of inflation-generated magnetic field through phase transitions}}, \href{https://doi.org/10.1103/PhysRevD.86.103005}{\emph{Phys. Rev. D} {\bfseries 86} (2012) 103005} [\href{https://arxiv.org/abs/1206.2428}{{\ttfamily 1206.2428}}].

\bibitem{Kisslinger:2015hua}
L.~Kisslinger and T.~Kahniashvili, \emph{{Polarized Gravitational Waves from Cosmological Phase Transitions}}, \href{https://doi.org/10.1103/PhysRevD.92.043006}{\emph{Phys. Rev. D} {\bfseries 92} (2015) 043006} [\href{https://arxiv.org/abs/1505.03680}{{\ttfamily 1505.03680}}].

\bibitem{Caprini:2009yp}
C.~Caprini, R.~Durrer and G.~Servant, \emph{{The stochastic gravitational wave background from turbulence and magnetic fields generated by a first-order phase transition}}, \href{https://doi.org/10.1088/1475-7516/2009/12/024}{\emph{JCAP} {\bfseries 12} (2009) 024} [\href{https://arxiv.org/abs/0909.0622}{{\ttfamily 0909.0622}}].

\bibitem{Binetruy:2012ze}
P.~Binetruy, A.~Bohe, C.~Caprini and J.-F.~Dufaux, \emph{{Cosmological Backgrounds of Gravitational Waves and eLISA/NGO: Phase Transitions, Cosmic Strings and Other Sources}}, \href{https://doi.org/10.1088/1475-7516/2012/06/027}{\emph{JCAP} {\bfseries 06} (2012) 027} [\href{https://arxiv.org/abs/1201.0983}{{\ttfamily 1201.0983}}].

\bibitem{Ellis:2018mja}
J.~Ellis, M.~Lewicki and J.M.~No, \emph{{On the Maximal Strength of a First-Order Electroweak Phase Transition and its Gravitational Wave Signal}}, \href{https://doi.org/10.1088/1475-7516/2019/04/003}{\emph{JCAP} {\bfseries 04} (2019) 003} [\href{https://arxiv.org/abs/1809.08242}{{\ttfamily 1809.08242}}].

\bibitem{Coleman:1977py}
S.~Coleman, \emph{{Fate of the false vacuum: Semiclassical theory}}, \href{https://doi.org/10.1103/PhysRevD.16.1248}{\emph{Phys. Rev. D} {\bfseries 15} (1977) 2929}.

\bibitem{Linde:1981zj}
A.D.~Linde, \emph{{Decay of the False Vacuum at Finite Temperature}}, \href{https://doi.org/10.1016/0550-3213(83)90072-X}{\emph{Nucl. Phys. B} {\bfseries 216} (1983) 421}.

\bibitem{LINDE198137}
A.D.~Linde, \emph{Fate of the false vacuum at finite temperature: Theory and applications}, \href{https://doi.org/10.1016/0370-2693(81)90281-1}{\emph{Phys. Lett. B} {\bfseries 100} (1981) 37}.

\bibitem{Ghiglieri:2015nfa}
J.~Ghiglieri and M.~Laine, \emph{{Gravitational wave background from Standard Model physics: Qualitative features}}, \href{https://doi.org/10.1088/1475-7516/2015/07/022}{\emph{JCAP} {\bfseries 07} (2015) 022} [\href{https://arxiv.org/abs/1504.02569}{{\ttfamily 1504.02569}}].

\bibitem{Ghiglieri:2020mhm}
J.~Ghiglieri, G.~Jackson, M.~Laine and Y.~Zhu, \emph{{Gravitational wave background from Standard Model physics: Complete leading order}}, \href{https://doi.org/10.1007/JHEP07(2020)092}{\emph{JHEP} {\bfseries 07} (2020) 092} [\href{https://arxiv.org/abs/2004.11392}{{\ttfamily 2004.11392}}].

\bibitem{Braaten:1989mz}
E.~Braaten and R.D.~Pisarski, \emph{{Soft Amplitudes in Hot Gauge Theories: A General Analysis}}, \href{https://doi.org/10.1016/0550-3213(90)90508-B}{\emph{Nucl. Phys. B} {\bfseries 337} (1990) 569}.

\bibitem{Weldon:1982bn}
H.A.~Weldon, \emph{{Effective Fermion Masses of Order gT in High Temperature Gauge Theories with Exact Chiral Invariance}}, \href{https://doi.org/10.1103/PhysRevD.26.2789}{\emph{Phys. Rev. D} {\bfseries 26} (1982) 2789}.

\bibitem{Fonseca:2020vke}
R.M.~Fonseca, \emph{{GroupMath: A Mathematica package for group theory calculations}}, \href{https://doi.org/10.1016/j.cpc.2021.108085}{\emph{Comput. Phys. Commun.} {\bfseries 267} (2021) 108085} [\href{https://arxiv.org/abs/2011.01764}{{\ttfamily 2011.01764}}].

\bibitem{Herman:2022fau}
N.~Herman, L.~Lehoucq and A.~F\'{u}zfa, \emph{{Electromagnetic antennas for the resonant detection of the stochastic gravitational wave background}}, \href{https://doi.org/10.1103/PhysRevD.108.124009}{\emph{Phys. Rev. D} {\bfseries 108} (2023) 124009} [\href{https://arxiv.org/abs/2203.15668}{{\ttfamily 2203.15668}}].

\bibitem{Gehrman:2023esa}
T.C.~Gehrman, B.~Shams Es~Haghi, K.~Sinha and T.~Xu, \emph{{The primordial black holes that disappeared: connections to dark matter and MHz-GHz gravitational Waves}}, \href{https://doi.org/10.1088/1475-7516/2023/10/001}{\emph{JCAP} {\bfseries 10} (2023) 001} [\href{https://arxiv.org/abs/2304.09194}{{\ttfamily 2304.09194}}].

\bibitem{Gatti:2024mde}
C.~Gatti, L.~Visinelli and M.~Zantedeschi, \emph{{Cavity detection of gravitational waves: Where do we stand?}}, \href{https://doi.org/10.1103/PhysRevD.110.023018}{\emph{Phys. Rev. D} {\bfseries 110} (2024) 023018} [\href{https://arxiv.org/abs/2403.18610}{{\ttfamily 2403.18610}}].

\bibitem{TitoDAgnolo:2024res}
R.~Tito~D'Agnolo and S.A.R.~Ellis, \emph{{Classical (and quantum) heuristics for gravitational wave detection}}, \href{https://doi.org/10.1007/JHEP04(2025)164}{\emph{JHEP} {\bfseries 04} (2025) 164} [\href{https://arxiv.org/abs/2412.17897}{{\ttfamily 2412.17897}}].

\bibitem{Espinosa:1996qw}
J.R.~Espinosa, \emph{{Dominant two loop corrections to the MSSM finite temperature effective potential}}, \href{https://doi.org/10.1016/0550-3213(96)00297-0}{\emph{Nucl. Phys. B} {\bfseries 475} (1996) 273} [\href{https://arxiv.org/abs/hep-ph/9604320}{{\ttfamily hep-ph/9604320}}].

\bibitem{Curtin:2016urg}
D.~Curtin, P.~Meade and H.~Ramani, \emph{{Thermal Resummation and Phase Transitions}}, \href{https://doi.org/10.1140/epjc/s10052-018-6268-0}{\emph{Eur. Phys. J. C} {\bfseries 78} (2018) 787} [\href{https://arxiv.org/abs/1612.00466}{{\ttfamily 1612.00466}}].

\bibitem{Niemi:2024vzw}
L.~Niemi and T.V.I.~Tenkanen, \emph{{Investigating two-loop effects for first-order electroweak phase transitions}}, \href{https://doi.org/10.1103/PhysRevD.111.075034}{\emph{Phys. Rev. D} {\bfseries 111} (2025) 075034} [\href{https://arxiv.org/abs/2408.15912}{{\ttfamily 2408.15912}}].

\bibitem{Antusch:2023zjk}
S.~Antusch, K.~Hinze, S.~Saad and J.~Steiner, \emph{{Singling out SO(10) GUT models using recent PTA results}}, \href{https://doi.org/10.1103/PhysRevD.108.095053}{\emph{Phys. Rev. D} {\bfseries 108} (2023) 095053} [\href{https://arxiv.org/abs/2307.04595}{{\ttfamily 2307.04595}}].

\bibitem{Buchmuller:2023aus}
W.~Buchmüller, V.~Domcke and K.~Schmitz, \emph{{Metastable cosmic strings}}, \href{https://doi.org/10.1088/1475-7516/2023/11/020}{\emph{JCAP} {\bfseries 11} (2023) 020} [\href{https://arxiv.org/abs/2307.04691}{{\ttfamily 2307.04691}}].

\bibitem{Lazarides:2023rqf}
G.~Lazarides, R.~Maji, A.~Moursy and Q.~Shafi, \emph{{Inflation, superheavy metastable strings and gravitational waves in non-supersymmetric flipped SU(5)}}, \href{https://doi.org/10.1088/1475-7516/2024/03/006}{\emph{JCAP} {\bfseries 03} (2024) 006} [\href{https://arxiv.org/abs/2308.07094}{{\ttfamily 2308.07094}}].

\bibitem{Fu:2023mdu}
B.~Fu, S.F.~King, L.~Marsili, S.~Pascoli, J.~Turner and Y.-L.~Zhou, \emph{{Testing realistic SO(10) SUSY GUTs with proton decay and gravitational waves}}, \href{https://doi.org/10.1103/PhysRevD.109.055025}{\emph{Phys. Rev. D} {\bfseries 109} (2024) 055025} [\href{https://arxiv.org/abs/2308.05799}{{\ttfamily 2308.05799}}].

\bibitem{Ahmed:2023rky}
W.~Ahmed, M.U.~Rehman and U.~Zubair, \emph{{Probing stochastic gravitational wave background from $SU(5)\times U(1)_\chi$ strings in light of NANOGrav 15-year data}}, \href{https://doi.org/10.1088/1475-7516/2024/01/049}{\emph{JCAP} {\bfseries 01} (2024) 049} [\href{https://arxiv.org/abs/2308.09125}{{\ttfamily 2308.09125}}].

\bibitem{King:2023wkm}
S.F.~King, G.K.~Leontaris and Y.-L.~Zhou, \emph{{Flipped SU(5): unification, proton decay, fermion masses and gravitational waves}}, \href{https://doi.org/10.1007/JHEP03(2024)006}{\emph{JHEP} {\bfseries 03} (2024) 006} [\href{https://arxiv.org/abs/2311.11857}{{\ttfamily 2311.11857}}].

\bibitem{Antusch:2024nqg}
S.~Antusch, K.~Hinze and S.~Saad, \emph{{Explaining PTA results by metastable cosmic strings from SO(10) GUT}}, \href{https://doi.org/10.1088/1475-7516/2024/10/007}{\emph{JCAP} {\bfseries 10} (2024) 007} [\href{https://arxiv.org/abs/2406.17014}{{\ttfamily 2406.17014}}].

\bibitem{Maji:2024tzg}
R.~Maji and Q.~Shafi, \emph{{Kinetic mixing, proton decay and gravitational waves in SO(10)}}, \href{https://doi.org/10.1007/JHEP10(2024)157}{\emph{JHEP} {\bfseries 10} (2024) 157} [\href{https://arxiv.org/abs/2408.14350}{{\ttfamily 2408.14350}}].

\bibitem{Maji:2025thf}
R.~Maji and Q.~Shafi, \emph{{Superheavy Metastable Strings in SO(10)}}, \href{https://doi.org/10.1007/JHEP06(2025)217}{\emph{JHEP} {\bfseries 06} (2025) 217} [\href{https://arxiv.org/abs/2504.09055}{{\ttfamily 2504.09055}}].

\bibitem{Ozer:2005dwq}
A.D.~\"Ozer, \emph{{$SO(10)$ - Grand Unification and Fermion Masses}}, Ph.D. thesis, Munich U., 2005.
\newblock 10.5282/edoc.4695.

\bibitem{PhysRevD.9.3320}
L.~Dolan and R.~Jackiw, \emph{Symmetry behavior at finite temperature}, \href{https://doi.org/10.1103/PhysRevD.9.3320}{\emph{Phys. Rev. D} {\bfseries 9} (1974) 3320}.

\bibitem{Ellis:2019oqb}
J.~Ellis, M.~Lewicki, J.M.~No and V.~Vaskonen, \emph{{Gravitational wave energy budget in strongly supercooled phase transitions}}, \href{https://doi.org/10.1088/1475-7516/2019/06/024}{\emph{JCAP} {\bfseries 06} (2019) 024} [\href{https://arxiv.org/abs/1903.09642}{{\ttfamily 1903.09642}}].

\bibitem{Caprini:2015zlo}
C.~Caprini et~al., \emph{{Science with the space-based interferometer eLISA. II: Gravitational waves from cosmological phase transitions}}, \href{https://doi.org/10.1088/1475-7516/2016/04/001}{\emph{JCAP} {\bfseries 04} (2016) 001} [\href{https://arxiv.org/abs/1512.06239}{{\ttfamily 1512.06239}}].

\bibitem{Caprini:2019egz}
C.~Caprini et~al., \emph{{Detecting gravitational waves from cosmological phase transitions with LISA: an update}}, \href{https://doi.org/10.1088/1475-7516/2020/03/024}{\emph{JCAP} {\bfseries 03} (2020) 024} [\href{https://arxiv.org/abs/1910.13125}{{\ttfamily 1910.13125}}].

\bibitem{Espinosa:2010hh}
J.R.~Espinosa, T.~Konstandin, J.M.~No and G.~Servant, \emph{{Energy Budget of Cosmological First-order Phase Transitions}}, \href{https://doi.org/10.1088/1475-7516/2010/06/028}{\emph{JCAP} {\bfseries 06} (2010) 028} [\href{https://arxiv.org/abs/1004.4187}{{\ttfamily 1004.4187}}].

\bibitem{PhysRevD.25.2074}
P.J.~Steinhardt, \emph{Relativistic detonation waves and bubble growth in false vacuum decay}, \href{https://doi.org/10.1103/PhysRevD.25.2074}{\emph{Phys. Rev. D} {\bfseries 25} (1982) 2074}.

\bibitem{Hindmarsh:2017gnf}
M.~Hindmarsh, S.J.~Huber, K.~Rummukainen and D.J.~Weir, \emph{{Shape of the acoustic gravitational wave power spectrum from a first order phase transition}}, \href{https://doi.org/10.1103/PhysRevD.96.103520}{\emph{Phys. Rev. D} {\bfseries 96} (2017) 103520} [\href{https://arxiv.org/abs/1704.05871}{{\ttfamily 1704.05871}}].

\bibitem{nakahara}
M.~Nakahara, \emph{Geometry, Topology and Physics}, Taylor \& Francis, 2nd~ed. (2003).

\bibitem{hall}
B.C.~Hall, \emph{Lie Groups, Lie Algebras, and Representations}, Springer, 2nd~ed. (2015).

\bibitem{humphreys}
J.E.~Humphreys, \emph{Introduction to Lie Algebras and Representation Theory}, Springer (1972).

\end{thebibliography}\endgroup

\end{document}